\documentclass[aps,prd,superscriptaddress,preprintnumbers,nofootinbib]{revtex4}
\usepackage{graphicx}
\usepackage{slashed}
\usepackage{verbatim}
\usepackage[colorlinks]{hyperref}
\usepackage{bm}
\usepackage{amsmath,amssymb}
\usepackage{mathtools}
\usepackage{float}
\usepackage{color}
\usepackage{leftidx}
\usepackage{booktabs}
\usepackage{latexsym}
\usepackage{revsymb}
\usepackage{multirow}
\usepackage{hypcap}
\usepackage{array}
\usepackage{tikz}
\hypersetup{
    colorlinks=red,
    citecolor=blue,
    filecolor=magenta,
    urlcolor=red,
}
\renewcommand{\d}{\mathrm{d}}
\newcommand{\be}{\begin{eqnarray}}
\newcommand{\ee}{\end{eqnarray}}
\newcommand{\pup}{p^\uparrow}

\newcommand{\bfk}{\mbox{\boldmath $k$}}
\newcommand*\circled[1]{\tikz[baseline=(char.base)]{
            \node[shape=circle,draw,inner sep=.0pt] (char) {#1};}}

\usepackage[normalem]{ulem}

\begin{document}

\title{Process dependence of the gluon Sivers function in $p^\uparrow p \to J/\psi + X$ \\
within a TMD approach in NRQCD}

\author{Umberto D'Alesio}
\email{umberto.dalesio@ca.infn.it}
\affiliation{Dipartimento di Fisica, Universit\`a di Cagliari, Cittadella Universitaria, I-09042 Monserrato (CA), Italy}
\affiliation{INFN, Sezione di Cagliari, Cittadella Universitaria, I-09042 Monserrato (CA), Italy}
\author{Luca Maxia}
\email{luca.maxia@ca.infn.it}
\affiliation{Dipartimento di Fisica, Universit\`a di Cagliari, Cittadella Universitaria, I-09042 Monserrato (CA), Italy}
\affiliation{INFN, Sezione di Cagliari, Cittadella Universitaria, I-09042 Monserrato (CA), Italy}

\author{Francesco Murgia}
\email{francesco.murgia@ca.infn.it}
\affiliation{INFN, Sezione di Cagliari, Cittadella Universitaria, I-09042 Monserrato (CA), Italy}

\author{Cristian Pisano}
\email{cristian.pisano@unica.it}
\affiliation{Dipartimento di Fisica, Universit\`a di Cagliari, Cittadella Universitaria, I-09042 Monserrato (CA), Italy}
\affiliation{INFN, Sezione di Cagliari, Cittadella Universitaria, I-09042 Monserrato (CA), Italy}

\author{Sangem Rajesh}
\email{rajesh.sangem@ca.infn.it}
\affiliation{Dipartimento di Fisica, Universit\`a di Cagliari, Cittadella Universitaria, I-09042 Monserrato (CA), Italy}
\affiliation{INFN, Sezione di Cagliari, Cittadella Universitaria, I-09042 Monserrato (CA), Italy}

\date{\today}
\begin{abstract}
We consider the transverse single-spin asymmetry (SSA) for $J/\psi$ production in $p^\uparrow p \to J/\psi +~X$ within a TMD approach in non-relativistic QCD. Extending a previous study~\cite{DAlesio:2019gnu}, we employ here the color-gauge invariant generalized parton model (CGI-GPM), in which spin and intrinsic transverse momentum effects are taken into account, together with leading-order initial- and final-state interactions (ISIs and FSIs). We find that, even when the heavy-quark pair is produced in a color-octet state, ISIs and FSIs lead to a nonvanishing SSA, allowing, in principle, to test the process dependence of the gluon Sivers function (GSF).
We show that of the two independent contributions, due to the so-called $f$- and $d$-type GSFs, appearing in the CGI-GPM, the $d$-type one turns out to be dynamically suppressed.
Therefore, as already found adopting the Color-Singlet Model approach for the $J/\psi$ formation~\cite{DAlesio:2017rzj}, only the $f$-type GSF could play a role in phenomenology. A comparison with the corresponding results obtained in the generalized parton model, without the inclusion of ISIs and FSIs, is also carried out.
\end{abstract}
\maketitle
\raggedbottom
\section{Introduction}
\label{sec1}

The study of the three-dimensional structure of hadrons is of fundamental importance for our comprehension of their properties. It has certainly reached a substantial level of accuracy, thanks to many efforts, carried out in the last two decades both theoretically and experimentally. Its understanding in terms of transverse momentum dependent parton distributions (TMDs) represents the main achievement in this context~\cite{Angeles-Martinez:2015sea,Aidala:2020mzt}. These functions, nonperturbative in nature, have been extracted from several fits to experimental data, coming from semi-inclusive deep inelastic scattering (SIDIS) and Drell-Yan (DY) processes. As a matter of fact, most of the information collected so far is mainly restricted to the quark sector, while gluon TMDs are still very poorly known.

Among the eight leading-twist nucleon TMDs, the Sivers function~\cite{Sivers:1989cc,Sivers:1990fh} plays a seminal role. It describes the asymmetric azimuthal distribution of unpolarized partons (quark and gluons) in a fast-moving transversely polarized nucleon and is related to the orbital motion of partons. It could be responsible for azimuthal and single-spin asymmetries (SSAs) in processes where one of the initial nucleons is transversely polarized w.r.t.~its direction of motion. Another important feature is its expected process dependence. This can be understood in terms of initial- and final-state interactions (ISIs and FSIs), encoded in Wilson lines (gauge links), essential to preserve gauge invariance. One of the most striking consequences is the expectation of a sign change between the Sivers function probed in SIDIS w.r.t.~the one probed in DY processes~\cite{Collins:2002kn}. This is usually referred to as  modified universality of the Sivers function.

For SIDIS and DY processes TMD factorization has been proven~\cite{Ji:2004xq,Collins:2011zzd,GarciaEchevarria:2011rb}, and their analyses are well consolidated, in contrast to more inclusive processes like $pp\to h + X$, where a TMD scheme, referred to as the generalized parton model (GPM) \cite{DAlesio:2007bjf}, is adopted as a phenomenological Ansatz. On the other hand, its success in describing many polarization observables and its role in looking for potential factorization breaking effects make it an important tool.

For these processes, the color gauge invariant extension of the generalized parton model (GPM), known as the CGI-GPM, has been developed in Refs.~\cite{Gamberg:2010tj,DAlesio:2011kkm} and further extended in Ref.~\cite{DAlesio:2017rzj}.
In this approach, ISI and FSI are taken into account assuming a single eikonal gluon exchange between the struck parton and the remnants of the transversely polarized proton. This approximation is basically the leading order contribution, in a perturbative expansion of the Wilson line, in the definition of the Sivers function.

As in the quark case, the process dependence of the gluon Sivers function (GSF) can be absorbed into the corresponding partonic hard functions entering the factorized expression of the cross sections. However, two universal, completely independent, Sivers distributions have to be introduced~\cite{Bomhof:2006ra,DAlesio:2017rzj}. The reason is that, for three colored gluons, there are two different ways of forming a color-singlet state: the totally antisymmetric color combination, even under charge conjugation, commonly referred to as an $f$-type state, and the symmetric combination, odd under $C$-parity, referred to as a $d$-type state.

Due mainly to the lack of experimental data, information on the GSF is very limited~\cite{Boer:2015vso}. Some initial attempts, within the GPM, have been made to constrain it from mid-rapidity data for  inclusive pion production at RHIC~\cite{Anselmino:2006yq,DAlesio:2015fwo,DAlesio:2018rnv}. A similar analysis has been also performed in Ref.~\cite{Godbole:2017syo}.

In order to probe the unknown GSF other processes have been considered, both in $ep$ and $pp$ collisions. Among them, the production of quarkonium states, like the $J/\psi$ meson, has been shown to have a great potential~\cite{Godbole:2012bx,Godbole:2014tha,Mukherjee:2016qxa,Boer:2016fqd,Rajesh:2018qks,Bacchetta:2018ivt, Zheng:2018awe,DAlesio:2017rzj,DAlesio:2018rnv,DAlesio:2019gnu}.

We notice here that the study of $J/\psi$ production is important by itself, and various models have been formulated to describe its formation mechanism. Among them, we recall here the Color-Singlet Model (CSM)~\cite{Berger:1980ni,Baier:1983va}, where the heavy-quark pair is directly produced with the same quantum numbers as the observed quarkonium state. A more rigorous theory was then developed, referred to as non-relativistic QCD (NRQCD), where the heavy-quark pair can be produced also in a color-octet state with different quantum numbers. This, subsequently, evolves into the physical quarkonium state by the nonperturbative emission of soft gluons. This framework implies a separation of short-distance coefficients, which can be calculated perturbatively as expansions in the strong-coupling constant $\alpha_s$, from long-distance matrix elements (LDMEs), to be extracted from experiment~\cite{Bodwin:1994jh}. These are predicted to scale with a definite power of the heavy-quark relative velocity $v$ in the limit $v\ll 1$. In this way, the theoretical predictions are organized as double expansions in $\alpha_s$ and $v$. For a detailed overview see Ref.~\cite{Lansberg:2019adr} and references therein.

A soft collinear effective field theory (SCET) approach to factorization for quarkonium production and decay, relevant for TMD extractions, has also appeared~\cite{Fleming:2019pzj,Echevarria:2019ynx}. Quite recently it has been also shown how one can obtain the proper matching between the high and low transverse momentum regime for $J/\psi$ production in SIDIS~\cite{Boer:2020bbd}.

As mentioned above, SSAs for quarkonium production in $pp$ collisions have been extensively studied in a series of papers by some of us, employing both the GPM and the CGI-GPM within the Color-Singlet Model~\cite{DAlesio:2017rzj}, and quite recently adopting the GPM within NRQCD~\cite{DAlesio:2019gnu}. Here we extend and complete these analyses, by adding an important piece of information and focusing on the role of ISIs and FSIs, i.e.~adopting the CGI-GPM approach, within NRQCD. We will discuss the interplay of ISIs and FSIs with the formation mechanism, carrying out a detailed study of their net effect when the quarkonium is produced in a color-singlet or in a color-octet state. A comparison with the results found in the simpler GPM approach, still within NRQCD, is also carried out.

It is worth noticing that we are mainly interested in SSAs in the region of small to moderate $P_T$ values for the $J/\psi$ (in the $pp$ center of mass frame), the TMD approach regime, while NRQCD is usually applied to moderate to large $P_T$ values. Special attention, following what discussed in detail in Ref.~\cite{DAlesio:2019gnu}, will be paid in this respect.

The paper is organized as follows: in Sec.~\ref{sec2} we summarize the formalism for the computation of single-spin asymmetries in $pp\to J/\psi + X$, adopting the CGI-GPM framework within NRQCD, which was presented in details in Refs.~\cite{DAlesio:2017rzj,DAlesio:2019gnu}, while results are shown and discussed in Sec.~\ref{sec3}. Conclusions and final remarks are gathered in Sec.~\ref{sec4}. All the expressions for the hard scattering amplitudes squared, computed within the CGI-GPM, are collected in Appendix~\ref{app:A}.

\section{Single-spin asymmetries in the CGI-GPM approach}
\label{sec2}
The SSA for the $p^\uparrow p\rightarrow J/\psi+X$ process is defined as follows:
\begin{equation}
\label{eq:AN}
A_N \equiv  \frac{\d \sigma^\uparrow - \d \sigma^\downarrow}{\d \sigma^\uparrow + \d \sigma^\downarrow} \equiv \frac{ \d\Delta\sigma}{ 2 \d\sigma}\,,
\end{equation}
where $\d \sigma^{\uparrow(\downarrow)}$ is the differential cross section, $E_h\,d^3\sigma^{\uparrow(\downarrow)}/d^3{\bm P}_h$, with one of the initial protons polarized along the transverse direction $\uparrow(\downarrow)$ with respect to the production plane. We consider the proton-proton collision along the $z$ axis in the center of mass frame, with the polarized proton moving along $+ \hat z$, wherein the $J/\psi$ is produced in the $x-z$ plane, and the $\uparrow$ transverse polarization is along $+\hat y$. The numerator of the asymmetry receives a sizeable contribution only from the Sivers function~\cite{DAlesio:2017rzj} which is defined as~\cite{Bacchetta:2004jz}
\be
\Delta \hat f_{a/\pup}(x_a, \bfk_{\perp a})  & \equiv &
\hat f_{a/\pup}(x_a, \bm{k}_{\perp a}) - \hat f_{a/p^\downarrow}(x_a, \bm{k}_{\perp a})\nonumber \\
\label{defsiv}
&= & \Delta^N f_{a/\pup}(x_a, k_{\perp a}) \, \cos\phi_a\nonumber \\
&= & -2 \, \frac{k_{\perp a}}{M_p} \, f_{1T}^{\perp a} (x_a, k_{\perp a}) \, \cos\phi_a \,.
\ee
This TMD describes the azimuthal distribution of an unpolarized parton $a$ with light-cone momentum fraction $x_a$ and intrinsic transverse momentum $\bm k_{\perp a}=k_{\perp a}(\cos\phi_a,\sin\phi_a,0)$ in a high-energy, transversely polarized nucleon with mass $M_p$, moving along the $z$ direction.

\begin{figure}[t]
\begin{center}
\includegraphics[height=3cm,width=17cm]{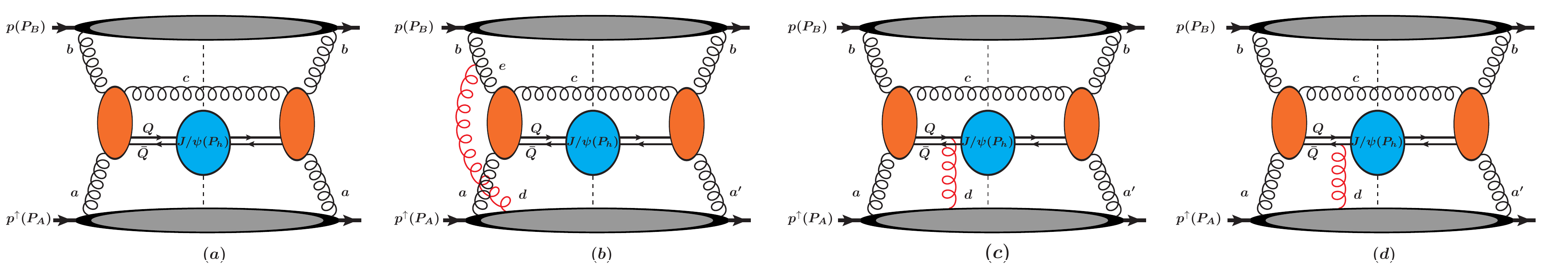}
\end{center}
\caption{Diagrams for the dominant gluon fusion contribution to the process $p^\uparrow p \to J/\psi + X$ in the GPM (a) and in the CGI-GPM approaches with inclusion, at leading order, of additional effects from initial-state (b) and final-state ((c) and (d)) interactions. FSIs are effective only when the $J/\psi$ is produced in a color-octet state. Notice that there are analogous diagrams for other $2\to 2$ subprocesses as well as for the $2\to 1$ channels, like $g+g\to J/\psi$.
The scattering amplitudes for the underlying partonic reaction, $g+g\to J/\psi +g$, are represented by the central blobs, while the upper and lower ones describe the soft proton~$\to$~gluon transitions.}
\label{fig:ppjpsi}
\end{figure}

In order to proceed with the calculation of the asymmetry within the CGI-GPM framework we take into account the proper insertion of the leading order contribution, in the strong coupling constant power expansion, of the gauge links, for all diagrams relevant for $J/\psi$ production in NRQCD (see, as an example, Fig.~\ref{fig:ppjpsi} for the gluon-gluon $2\to 2$ channel). We note that we did not consider the FSIs of the unobserved particle (gluon) because they are known to vanish after summing the different cut diagrams, see for example the discussion in Ref.~\cite{Gamberg:2010tj}.

One has to include the $2\rightarrow 1$ partonic subprocesses, namely $g+g\rightarrow J/\psi$ and $q+\bar{q}\rightarrow J/\psi$, as well as the $2\rightarrow 2$ ones, that is $g+g\rightarrow J/\psi+g$, $g+q(\bar q)\rightarrow J/\psi+q(\bar q)$ and $q+\bar{q}\rightarrow J/\psi+g$, and exploit the contributions from the $\leftidx{^{3}}{S}{_1}^{(1,8)}$, $\leftidx{^{1}}{S}{_0}^{(8)}$ and $\leftidx{^{3}}{P}{_J}^{(8)}$ states respectively. Here we refer to the standard notation for a heavy-quark pair state $^{2S+1}L_J^{(c)}$, where $S$ is  the spin of the pair, $L$ and $J$ the orbital and total angular momentum and $c$ the color configuration, with $c=1, 8$. Notice that since in the $2 \to 1$ channels, at leading order, the $J/\psi$ gets its transverse momentum only from the intrinsic ones of the two initial partons, these contributions can be relevant only in the low-$P_T$ region.

As one can see, in the framework of NRQCD we have to take into account contributions to the SSA coming also from the quark Sivers function, not present in the CSM, where only the gluon-gluon fusion channel is at work. Moreover, as already pointed out previously, in the CGI-GPM formalism we have to consider two possible independent sources for the GSF.

Following Refs.~\cite{DAlesio:2019gnu,DAlesio:2017rzj}, to which we refer the reader for more details, the numerator of the asymmetry for $2\rightarrow 1$ channels, $a+b\to J/\psi$, i.e.~$g+g\to J/\psi$ and $q+\bar q \to J/\psi$, within the CGI-GPM approach is given by
\be
\label{eq:A2d1}
\d\Delta\sigma^{\mathrm{CGI-GPM}}_{2\to 1}
& = &\frac{2\pi}{x_a x_b s^2}\int d^2{\bm k}_{\perp a}d^2{\bm k}_{\perp b} \,\delta^2 ({\bm k}_{\perp a} + {\bm k}_{\perp b} - \bm{P}_T) \nonumber\\
& &\mbox{} \times \Big(- \frac{k_{\perp a}}{M_p}\Big) \cos\phi_a
\Big\{\sum_{q} \Big[ f_{1T}^{\perp q} (x_a, k_{\perp a}) \,f_{\bar q/p}(x_b, k_{\perp b})\, |\mathcal{M}^{\mathrm{Inc}}_{q\bar q\to J/\psi}|^2\Big] \nonumber \\
&& + \sum_{m=f,d} f_{1T}^{\perp g(m)} (x_a, k_{\perp a})\, f_{g/p}(x_b, k_{\perp b}) \,|\mathcal{M}^{\mathrm{Inc}(m)}_{g g\to J/\psi}|^2 \Big\}\,,
\ee
where  $q =u, d, s, \bar u, \bar d, \bar s$, and $\cal{M}^{\rm inc}$ are the hard scattering amplitudes modified by ISIs and FSIs, as detailed in the sequel. Moreover, at order $O(k_\perp/\sqrt{s})$,
\be
x_a = 
\frac{M_T}{\sqrt{s}}\, e^y \,,\hspace*{1cm}
x_b = 
\frac{M_T}{\sqrt{s}}\, e^{-y},
\ee
with $M_T = \sqrt {\bm{P}_T^2+M^2}$, being $M$ and $\bm{P}_T$ the mass of the $J/\psi$ and its transverse momentum, respectively, and $y$ its rapidity.

Similarly for $2\rightarrow 2$ channels, $a+b\to J/\psi + c$ (i.e.~$g+g\rightarrow J/\psi+g$, $g+q(\bar q)\rightarrow J/\psi+q(\bar q)$ and $q+\bar{q}\rightarrow J/\psi+g$), the numerator in Eq.~\eqref{eq:AN} is given by
\be
\d\Delta\sigma^{\mathrm{CGI-GPM}}_{2\to 2}
& = &\frac{1}{(2\pi)^2}\frac{1}{2s} \int \frac{dx_a}{x_a} \frac{dx_b}{x_b}\,
d^2{\bm k}_{\perp a}\,d^2{\bm k}_{\perp b} \, \delta(\hat{s}+\hat{t}+\hat{u}-M^2) \Big(-\frac{k_{\perp a}}{M_p}\Big) \cos\phi_a \nonumber\\
& & \mbox{} \times
\Big\{\sum_{q} \Big[ f_{1T}^{\perp q} (x_a, k_{\perp a}) \,\Big( f_{\bar q/p}(x_b, k_{\perp b})\, |\mathcal{M}^{\mathrm{Inc}}_{q\bar q\to J/\psi +g}|^2
+ f_{g/p}(x_b, k_{\perp b})\, |\mathcal{M}^{\mathrm{Inc}}_{qg\to J/\psi+q}|^2\Big) \Big] \nonumber\\
&& \mbox{} + \sum_{m=f,d}f_{1T}^{\perp g(m)} (x_a, k_{\perp a}) \Big( \sum_q f_{q/p}(x_b, k_{\perp b})\, |\mathcal{M}^{\mathrm{Inc} (m)}_{gq\to J/\psi+q}|^2
 + f_{g/p}(x_b, k_{\perp b})\,|\mathcal{M}^{\mathrm{Inc}(m)}_{g g\to J/\psi+g}|^2\Big) \Big\}\,.
\label{eq:N2to2}
\ee
The denominator in Eq.~\eqref{eq:AN} is just twice the unpolarized cross section, discussed in detail within a TMD scheme in Ref.~\cite{DAlesio:2019gnu}. For completeness we give the expressions separately for the $2\to 1$ and the $2\to 2$ channels:
\begin{equation}
\label{2d1bis}
E_h\frac{d^3\sigma^{2\to 1}}{d^3{\bm P}_h}=\sum_{a,b}\frac{\pi}{x_a x_b s^2}\int d^2{\bm k}_{\perp a}d^2{\bm k}_{\perp b}
f_{a/p}(x_a, k_{\perp a})f_{b/p}(x_b, k_{\perp b})\,\delta^2 ({\bm k}_{\perp a} + {\bm k}_{\perp b} - \bm{P}_T) |\mathcal{M}_{ab\rightarrow J/\psi }|^2\,,
\end{equation}
\begin{equation}\label{d2}
E_h\frac{d^3\sigma^{2\to 2}}{d^3{\bm P}_h}=
\frac{1}{2(2\pi)^2}\frac{1}{2s}\sum_{a,b,c}\int \frac{dx_a}{x_a} \frac{dx_b}{x_b}\,  d^2{\bm k}_{\perp
a}d^2{\bm k}_{\perp b} f_{a/p}(x_a,k_{\perp a})f_{b/p}(x_b,k_{\perp b})\delta(\hat{s}+\hat{t}+\hat{u}-M^2)
|\mathcal{M}_{ab\rightarrow J/\psi + c}|^2\,.
\end{equation}

All the hard scattering amplitudes squared $|\mathcal{M}^{\mathrm{Inc}}|^2$, where we have omitted the final state quantum numbers, are calculated perturbatively by incorporating, at leading order, the ISIs and FSIs within the CGI-GPM approach, and are listed in Appendix~\ref{app:A}, for all color octet states. The expressions for the color-singlet contributions can be directly found in Ref.~\cite{DAlesio:2017rzj}, where SSAs within the Color-Singlet Model were discussed.

Here we only point out that, for the $2\to 1$ processes, due to cancellations between the ISIs and FSIs, $|\mathcal{M}^{\mathrm{Inc}(f,d)}|^2=0$ for the $ g+g\rightarrow J/\psi$ subprocess independently of the $^{2S+1}L_J^{(c)}$ state, leaving active only the $q\bar q$ channel (second line in Eq.~(\ref{eq:A2d1})).
Moreover, for the $g+g\rightarrow J/\psi+g$ subprocess, the hard parts corresponding to $f_{1T}^{\perp g(d)}$, $|\mathcal{M}^{\mathrm{Inc}(d)}|^2$, turn out to be zero for all states (see also Ref.~\cite{Yuan:2008vn}). This means that, as already found in the study of SSAs for $J/\psi$ production within the CSM~\cite{DAlesio:2017rzj}, only the $f$-type GSF enters the dominant $gg$ channel. As we will see, this has important consequences in the phenomenological study.

\section{Numerical Results}
\label{sec3}

We proceed now with the phenomenological analysis of SSAs for $J/\psi$ production within the CGI-GPM approach. To this aim, following Refs.~\cite{DAlesio:2015fwo,DAlesio:2018rnv}, we adopt for the unpolarized TMDs a Gaussian factorized form
\be
f_{a/p}(x_a,k_{\perp a}) = \frac{e^{-k_{\perp a}^2/\langle k_{\perp a}^2 \rangle}}{\pi \langle k_{\perp a}^2\rangle} f_{a/p}(x_a)\,,
\ee
where $f_{a/p}(x_a)$ is the collinear parton distribution. The Sivers function is parameterized as
\begin{equation}
\Delta^N\! f_{a/p^\uparrow}(x_a,k_{\perp a}) =   \left (-2\frac{k_{\perp a}}{M_p}  \right )f_{1T}^{\perp\,a}
(x_a,k_{\perp a})  = 2 \, {\cal N}_a(x_a)\,f_{a/p}(x_a)\,
h(k_{\perp a})\,\frac{e^{-k_{\perp a}^2/\langle k_{\perp a}^2 \rangle}}
{\pi \langle k_{\perp a}^2 \rangle}\,,
\label{eq:siv-par-1}
\end{equation}
 with
\begin{equation}
{\cal N}_a(x_a) = N_a x_a^{\alpha}(1-x_a)^{\beta}\,
\frac{(\alpha+\beta)^{(\alpha+\beta)}}
{\alpha^{\alpha}\beta^{\beta}}\,,
\label{eq:nq-coll}
\end{equation}
where $|N_a|\leq 1$ and
\begin{equation}
h(k_{\perp a}) = \sqrt{2e}\,\frac{k_{\perp a}}{M'}\,e^{-k_{\perp a}^2/M'^2}\,.
\label{eq:h-siv}
\end{equation}
Eq.~(\ref{eq:siv-par-1}) can be rewritten as
\begin{equation}
\label{eq:siv-par}
\Delta^N\! f_{a/p^\uparrow}(x_a,k_{\perp a}) =   \frac{\sqrt{2e}}{\pi}   \,2\, {\cal N}_a(x_a)\, f_{a/p}(x_a) \, \sqrt{\frac{1-\rho_a}{\rho_a}}\,k_{\perp a}\, \frac{e^{-k_{\perp a}^2/ \rho_a\langle k_{\perp a}^2 \rangle}}
{\langle k_{\perp a}^2 \rangle^{3/2}}\,,
\end{equation}
where $\rho_a = \frac{M'^2}{\langle k_{\perp a}^2 \rangle +M'^2}$ and $0< \rho_a < 1$. With these choices, the Sivers function satisfies the model independent positivity bound for all values of $x_a$ and $k_{\perp a}$:
\begin{equation}
\vert \Delta^N f_{a/\pup}\,(x_a, k_{\perp a}) \vert  \le 2\,f_{a/p}\,(x_a, k_{\perp a})
\,,~~{\mathrm{ or}}~~
\frac{k_{\perp a}}{M_p}\, \vert f_{1T}^{\perp a} (x_a, k_{\perp a})\vert \le  f_{a/p}\,(x_a, k_{\perp a})~.
\end{equation}

For the collinear unpolarized parton distributions we will adopt the CTEQL1 set~\cite{Pumplin:2002vw}, at the factorization scale equal to $M_T$, adopting the DGLAP evolution equations.

These parameterizations allow us to integrate analytically the expressions entering the numerator and the denominator of $A_N$ for the $2\to 1$ channels, as follows:
\be
2d\sigma^{2\to 1} = \frac{1}{s^2}\sum_{a,b}\frac{1}{x_a x_b} \frac{1}{\langle k_{\perp a}^2 \rangle + \langle k_{\perp b}^2 \rangle} \exp\!\bigg(\!-\frac{P_T^2}{\langle k_{\perp a}^2 \rangle + \langle k_{\perp b}^2 \rangle}\bigg)  2\,f_{a/{p}}(x_a)\,f_{b/p}(x_b)\, |\mathcal{M}_{ab\to J/\psi}|^2,
\label{eq:D2to1}
\ee
\be
 d\Delta\sigma^{2\to 1} &=& \frac{\sqrt{2e}}{s^2}\sum_{a,b}\frac{1}{x_a x_b}\,
 \frac{\sqrt{\rho_a^3 (1-\rho_a)\langle k_{\perp a}^2\rangle}}{(\rho_a\langle k_{\perp a}^2 \rangle + \langle k_{\perp b}^2 \rangle)^2} P_T \exp\!\bigg(\!-\frac{P_T^2}{\rho_a\langle k_{\perp a}^2 \rangle + \langle k_{\perp b}^2 \rangle}\bigg) 2 \, {\cal N}_a(x_a)\, f_{a/{p}}(x_a)\,f_{b/p}(x_b)\, |\mathcal{M}_{ab\to J/\psi}|^2\,,\nonumber\\
 &&
\label{eq:N2to1}
\ee
where, while in Eq.~(\ref{eq:D2to1}) $(a,b) = (q,\bar q), (g,g)$, in Eq.~(\ref{eq:N2to1}) only the $q\bar q$ channel is active. For the $2\to 2$ channels we will have to proceed by a numerical integration.

At this point, we have to fix all parameters entering our expressions. In order to carry out a direct and easier comparison with the corresponding analysis performed in the GPM framework and study the impact of ISIs and FSIs, we adopt the same choices made in Ref.~\cite{DAlesio:2019gnu}. More precisely, for the quark unpolarized Gaussian width we use $\langle k_{\perp q}^2 \rangle= 0.25$~GeV$^2$, as extracted in Ref.~\cite{Anselmino:2005nn}, while for the gluon TMD we use $\langle k_{\perp g}^2 \rangle= 1$ GeV$^2$, that allows for a reasonably good description of the unpolarized cross section data in the low-$P_T$ region relevant for our study~\cite{DAlesio:2017rzj,DAlesio:2019gnu}.

Moving to the LDMEs, we consider the BK11~\cite{Butenschoen:2011yh} and the SYY13~\cite{Sun:2012vc} sets, whose values are given in Appendix~\ref{app:A}. As extensively discussed in Ref.~\cite{DAlesio:2019gnu} (where all details and physics motivations can be found), these sets are suitable enough for a study, within a TMD framework, of the low-$P_T$ region, where also SSA data are available: see, for instance, Figs.~1 and 2, left panels, in Ref.~\cite{DAlesio:2019gnu}, for a comparison with unpolarized cross section data.

We will start focusing on the relevance of ISIs and FSIs, and their interplay with the production mechanism, by a detailed comparison with the corresponding estimates in the GPM approach. To properly study the role of the partonic dynamics, we compute the contributions to $A_N$ by maximizing, separately, the Sivers effect both for quarks and gluons. This can be obtained by using $\rho_{q,g}=2/3$, ${\cal N}_{q}(x)=+1$ and ${\cal N}_{g}^{(f,d)}(x)=+1$ in Eq.~(\ref{eq:siv-par}). Notice that the chosen positive normalization allows for a better understanding of the relative signs coming from the hard dynamics.
Recalling that in NRQCD the heavy-quark pair can be produced in a CS or a CO state, and the latter with different angular momentum quantum numbers, we will also discuss separately each contribution. In the following, we will adopt the kinematics of the PHENIX experiment, at $\sqrt s= 200$ GeV, for which SSA data are available.

As we will show below, and as it happens in the GPM, also in the CGI-GPM the quark-initiated contributions are negligible. Similarly, the contribution from the $d$-type GSF is extremely small, due to the absence of the $gg$ channel in the numerator of the SSA (see comment at the end of the previous section). This means that within the CGI-GPM and NRQCD frameworks, one can still concentrate on the $f$-type gluon Sivers function.

\begin{figure}[t]
\begin{center}
\includegraphics[trim = 1.cm 0cm 1cm 0cm, width=8.5cm]{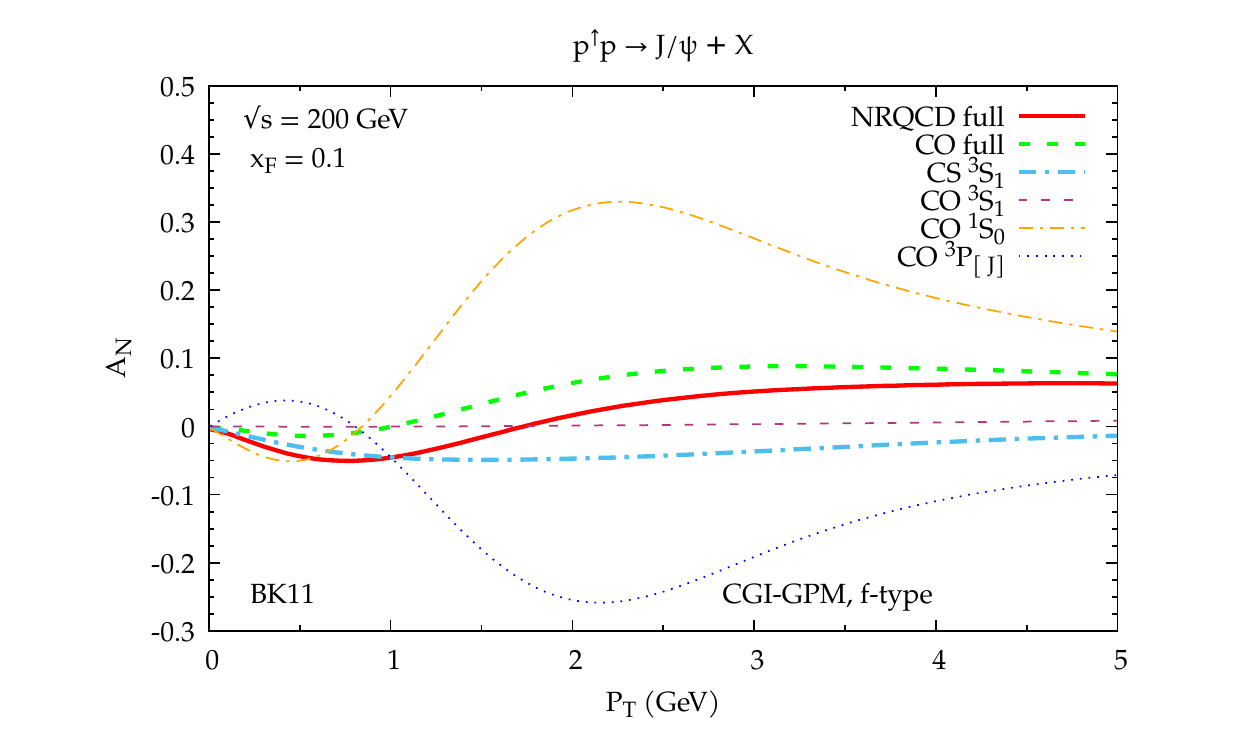}
\includegraphics[trim = 1.cm 0cm 1cm 0cm, width=8.5cm]{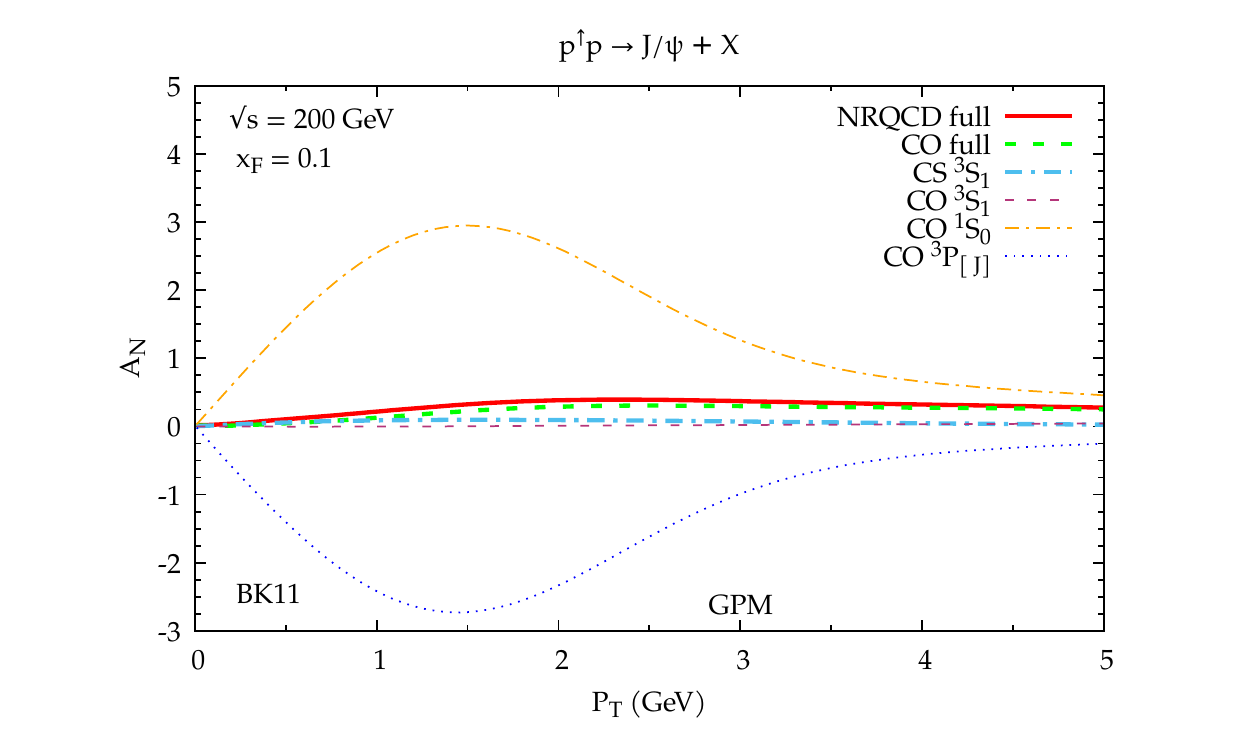}
\caption{Maximized contributions from the GSF to $A_N$ as a function of $P_T$ for the process
$p^\uparrow p\to J/\psi + X$ at $\sqrt s=200$ GeV and $x_F=0.1$  within the CGI-GPM, $f$-type GSF, (left panel) and the GPM (right panel) approaches by taking $\mathcal{N}_g(x)=+1$, $\rho_g=2/3$, and adopting the BK11 LDME set. The full result (red solid lines) together with its wave decomposition (see legend) are shown.
}
\label{fig:GPMvsCGI-waves}
\end{center}
\end{figure}

In Fig.~\ref{fig:GPMvsCGI-waves} we show our maximized estimates for $A_N$ at $\sqrt s=200$ GeV, coming from the dominant gluon contribution (red solid lines) in the CGI-GPM, $f$-type, (left panel) and the GPM (right panel) approaches, at $x_F=0.1$ as a function of $P_T$, together with a full wave decomposition (see the legend). Note the one order of magnitude difference in the vertical scale between the two panels. We also observe that some GPM contributions are larger than one since the denominator of the SSA includes all terms (entering with relative signs), while in the numerator we consider, term by term, only a specific wave state. The overall result (red solid lines) is, as it has to be, smaller than one.

We can further observe that, while the CGI-GPM estimates show clear oscillations, with a change of sign around $P_T\simeq 1$~GeV for CO states, the corresponding ones within the GPM have all a definite sign. This oscillating behaviour is specifically due to the $2\to 2$ channels and plays a role also in the GPM.

On the other hand, in the GPM the $2\to 1$ channels, at least in the small-$P_T$ region where they are relevant, compensate for this effect, leading to an overall definite sign. In contrast, as already pointed out in the previous section, in the CGI-GPM the gluon contributions from the $2\to 1$ channels are identically zero.
Notice that this oscillation in sign for the $2\to 2$ channels has nothing to do with the role of ISIs and FSIs (as it is clear from the fact that also the GPM estimates present this feature) and comes \emph{directly} from the parton dynamics, as weighted by the Sivers azimuthal phase. This can be easily understood recalling that within the GPM the hard parts in the numerator are identical to those in the denominator, which does not manifest any oscillation in sign. We also observe (Fig.~\ref{fig:GPMvsCGI-waves}) that this oscillating behaviour does not affect CS states, whose amplitudes squared present a much simpler structure in terms of their Mandelstam invariant dependence.

Another interesting feature in Fig.~\ref{fig:GPMvsCGI-waves} is that among the CO contributions, two of them, the $^1S_0$ and $^3P_{[J]}$ (where the symbol $[J]$ stands for a sum over $J=0,1,2$) wave terms, are very large, almost comparable in size but opposite in sign (this is due to the sign of the corresponding LDMEs), while one, the $^3S_1$ wave, is extremely small. This happens in both approaches. Moreover, the CS contribution, which as already said has a definite sign, shifts the zero in the CGI-GPM estimates to larger $P_T$ values, as compared with the CO terms.

Coming back to the size of each contribution, the smaller values in the CGI-GPM approach w.r.t.~the corresponding ones in the GPM are directly due to the cancellations between different hard partonic parts, entering with proper color factors and having in some cases opposite signs.

\begin{figure}[t]
\begin{center}
\includegraphics[trim = 1.cm 0cm 1cm 0cm, width=8.5cm]{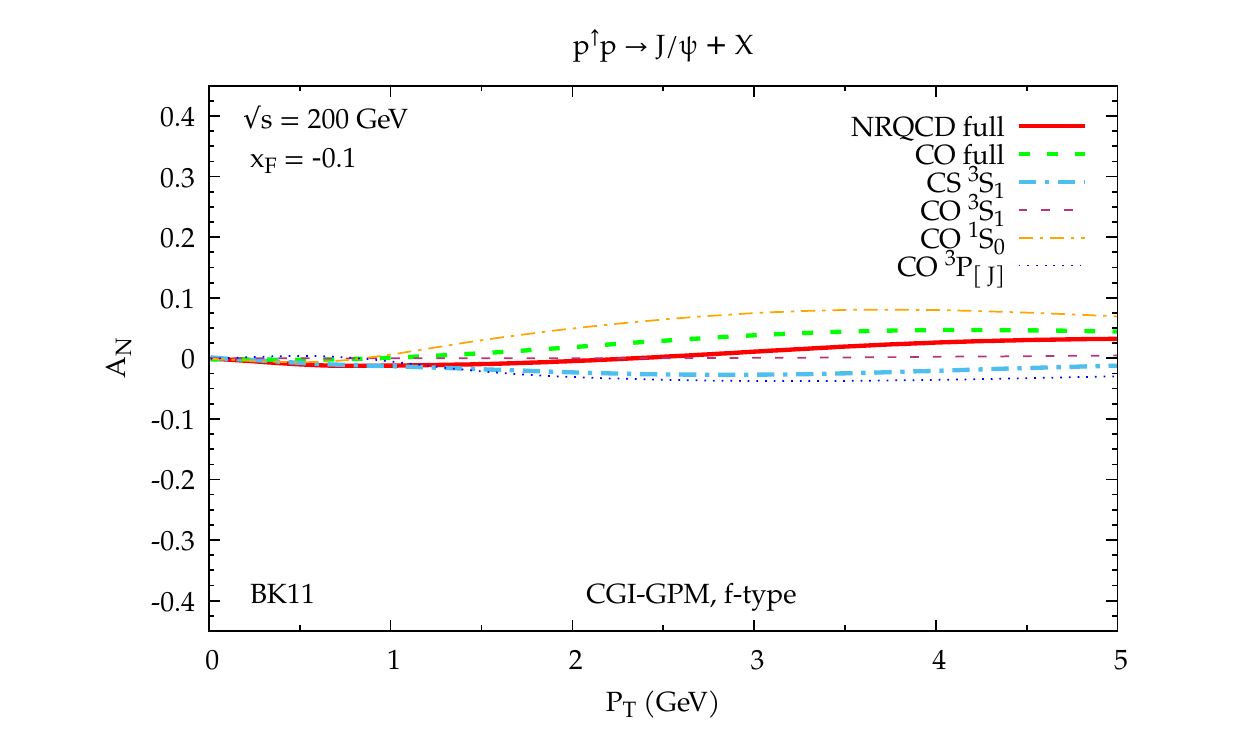}
\includegraphics[trim = 1.cm 0cm 1cm 0cm, width=8.5cm]{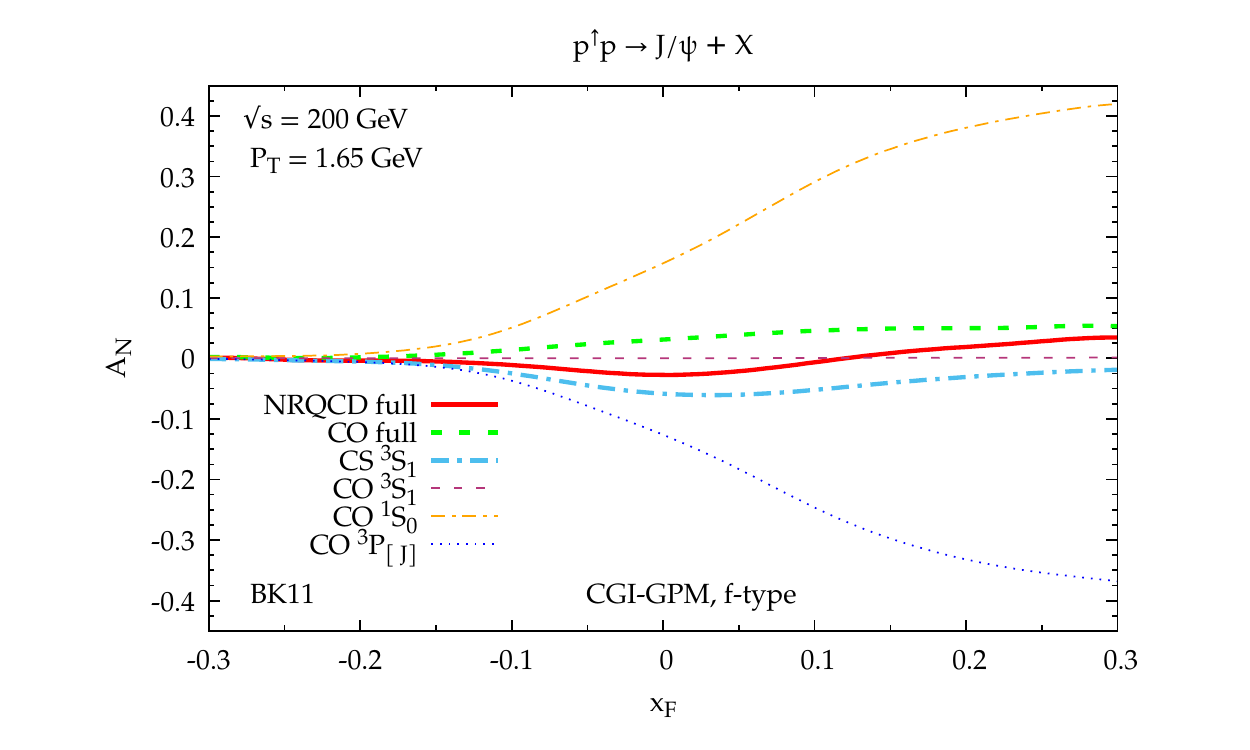}
\end{center}
\caption{Maximized contributions from the $f$-type GSF to $A_N$ for the process $p^\uparrow p\to J/\psi + X$ at $\sqrt s=200$ GeV as a function of $P_T$ at $x_F=-0.1$ (left panel) and as a function of $x_F$ at $P_T=1.65$ GeV (right panel) within the CGI-GPM approach by taking $\mathcal{N}_g(x)=+1$, $\rho_g=2/3$, and adopting the BK11 LDME set. The full result (red solid lines) together with its wave decomposition (see legend) are shown.
}
\label{fig:CGI-waves2}
\end{figure}

In Fig.~\ref{fig:CGI-waves2} we show the corresponding maximized estimates at $\sqrt s= 200$ GeV, within the CGI-GPM, at $x_F=-0.1$ as a function of $P_T$ (left panel) and at fixed $P_T=1.65$ GeV as a function of $x_F$ (right panel), as in the PHENIX analysis (see below).
Concerning the backward region, the main aspect is the suppression of all contributions as compared to those at $x_F=0.1$ (Fig.~\ref{fig:GPMvsCGI-waves}, left panel), that leads to much smaller results. In fact, besides the effects already discussed, the dependence on the Sivers azimuthal phase, $\cos\phi_a$ (through the Mandelstam invariants, see Appendix~\ref{app:A}), is less relevant in the hard parts, and therefore the integration over it (see Eq.~(\ref{eq:N2to2})) is more effective in reducing the effect.
Moreover, at such fixed $P_T$ value (right panel), as it is evident from Fig.~\ref{fig:GPMvsCGI-waves}, the cancellation among the various contributions in the CGI-GPM is much more effective for all $x_F$ values, leading to very small \emph{maximized} SSAs. The use of the other LDME set (SYY13) gives very similar results.

\begin{figure}[t]
\begin{center}
\includegraphics[trim = 1.cm 0cm 1cm 0cm, width=8.5cm]{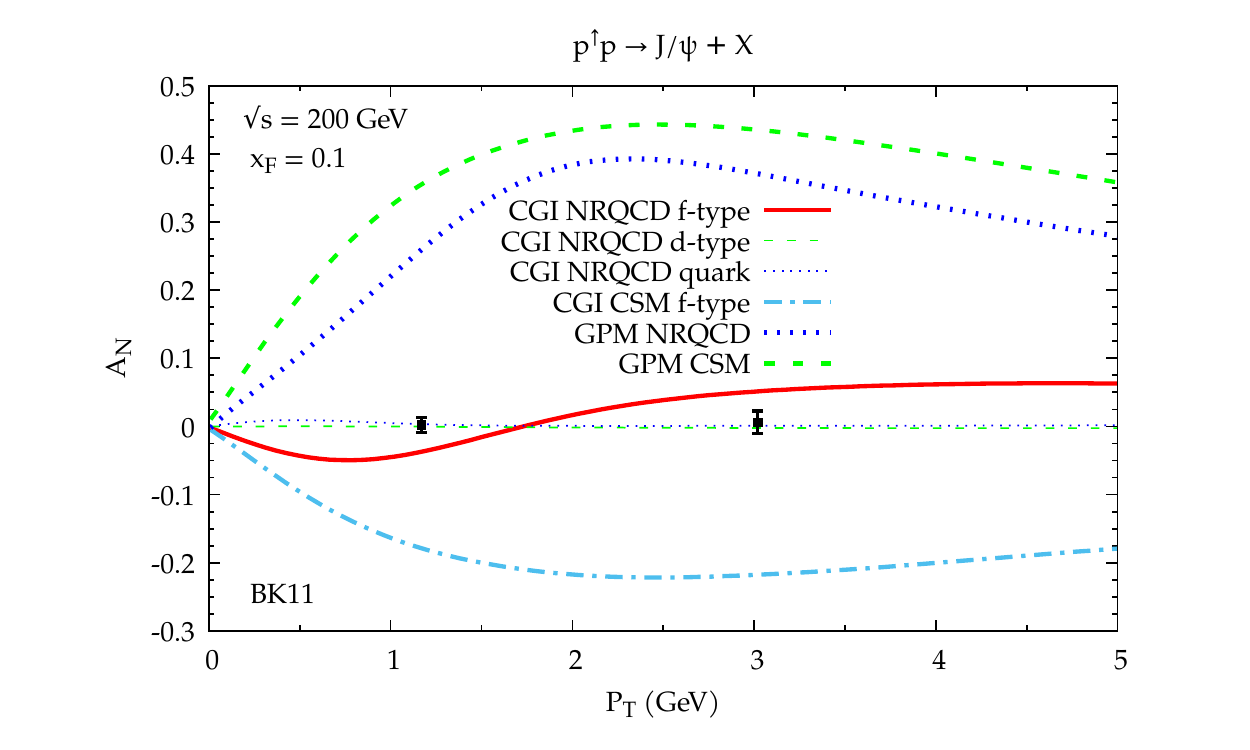}
\includegraphics[trim = 1.cm 0cm 1cm 0cm, width=8.5cm]{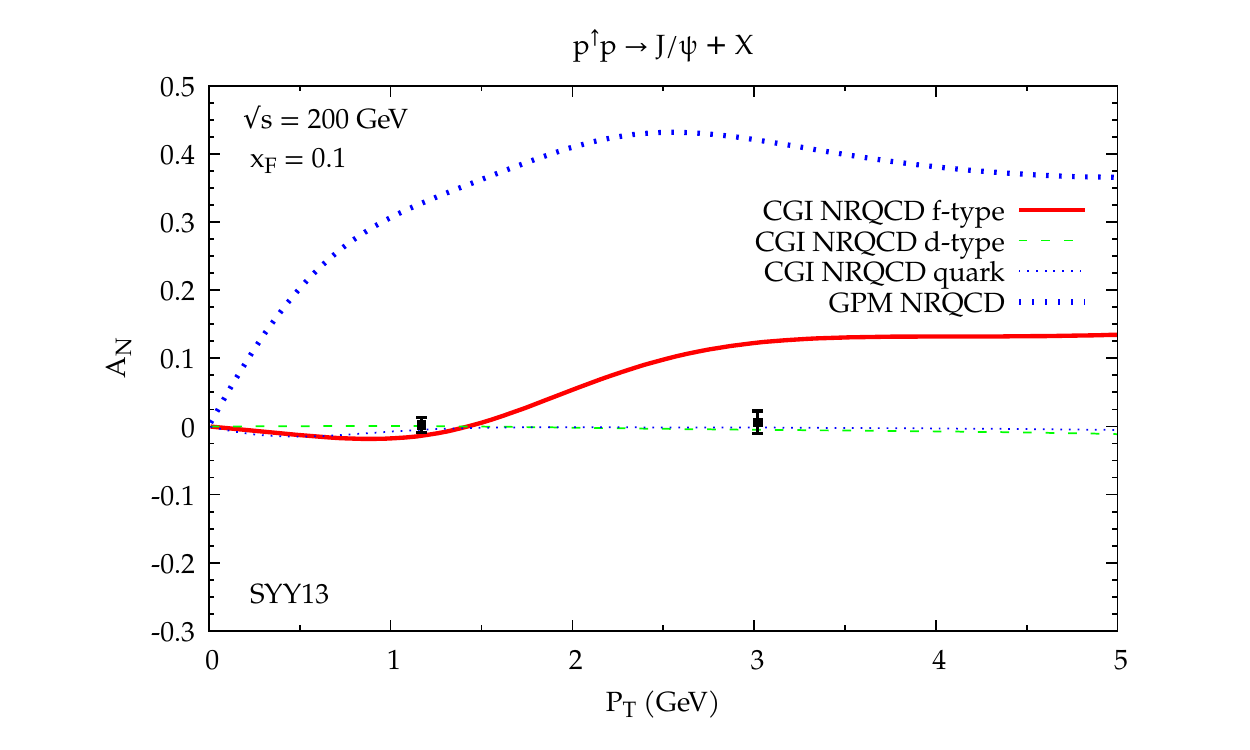}
\end{center}
\caption{Maximized $A_N$ estimates as a function of $P_T$ for the process
$p^\uparrow p\to J/\psi + X$ at $\sqrt s=200$ GeV and $x_F=0.1$  adopting the CGI-GPM and GPM approaches, within the CS model and NRQCD for the BK11 (left panel) and the SYY13 (right panel) LDME sets. Data are taken from \cite{Aidala:2018gmp}.}
\label{fig:allxF01}
\end{figure}

Having analysed in Figs. \ref{fig:GPMvsCGI-waves} and \ref{fig:CGI-waves2} the maximized contribution of the $f$-type GSF, to have a more complete view, in Fig.~\ref{fig:allxF01} we show for the BK11 (left panel) and the SYY13 (right panel) LDME sets a collection of results for maximized $A_N$ at $\sqrt s= 200$ GeV and $x_F=0.1$ as a function of $P_T$, adopting different models and approaches. PHENIX data~\cite{Aidala:2018gmp} are also shown. More precisely, we present the maximized estimates obtained within NRQCD, employing both the CGI-GPM (this work) and the GPM~\cite{DAlesio:2019gnu}, and those within the CSM in both approaches~\cite{DAlesio:2017rzj}. For the CGI-GPM all contributions are shown and, as already pointed out above, those from the quark (with the exception of the very small $P_T$ region for the SYY13 set) and the $d$-type gluon Sivers functions are negligible.
We also recall that the SYY13 LDME set includes only color-octet states. This implies, still within the CGI-GPM, larger values of $A_N$ at large $P_T$, due to the missing of the relative cancellation between the CO and the CS contributions, at work for the BK11 set. Notice that the estimates obtained within the CSM do not depend on the LDME set and, in this respect, they could appear identical also in the right panel.
From these plots we see that, even if to a much lesser extent as compared to the GPM, also within CGI-GPM one can potentially put some constraints on the size of the $f$-type GSF already with the few data points available.

\begin{figure}[H]
\begin{center}
\includegraphics[trim = 1.cm 0cm 1cm 0cm, width=8.5cm]{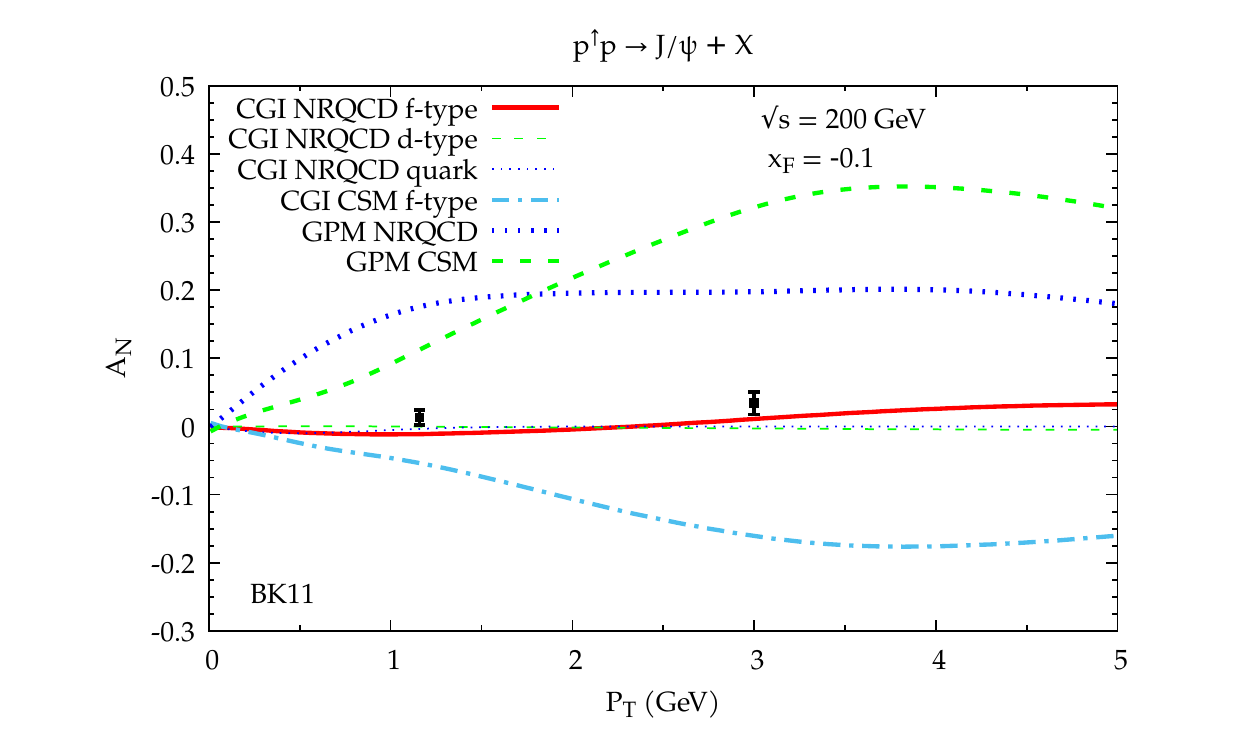}
\includegraphics[trim = 1.cm 0cm 1cm 0cm, width=8.5cm]{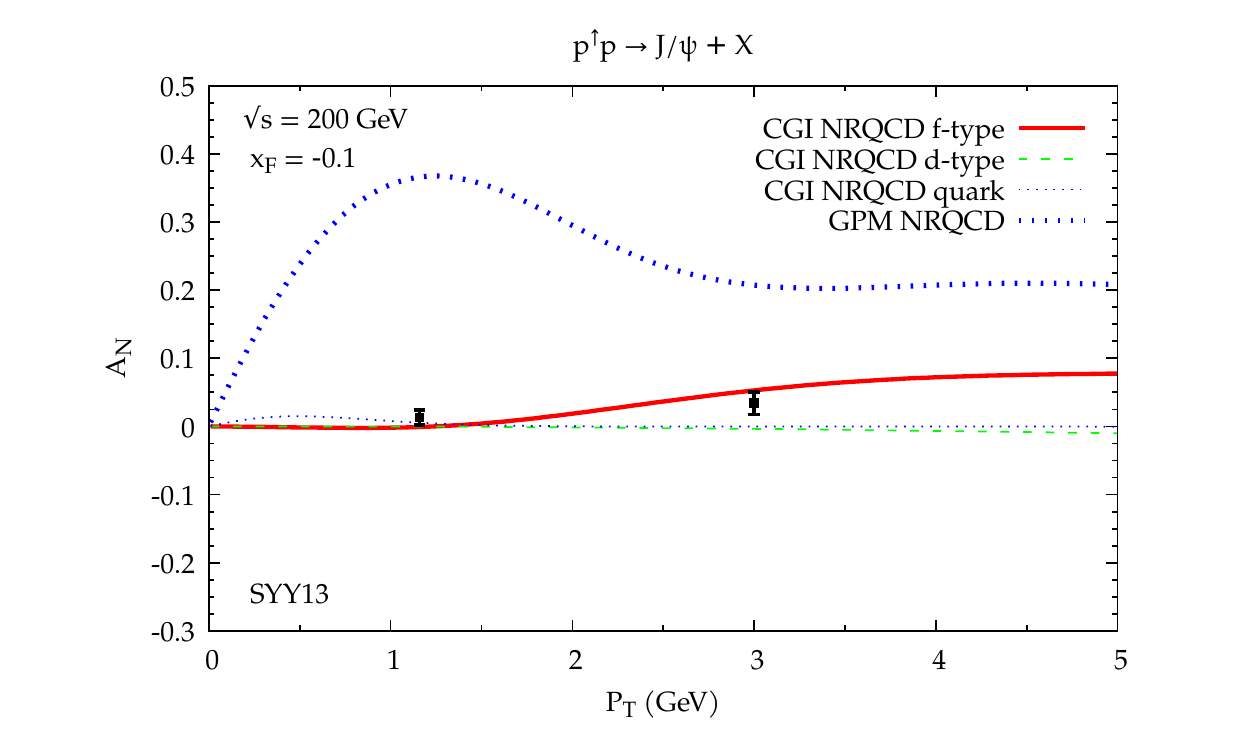}
\end{center}
\caption{Same as in Fig.~\ref{fig:allxF01} but for $x_F=-0.1$.}
\label{fig:allxF-01}
\end{figure}

As expected from previous considerations, the corresponding results at $x_F=-0.1$ show a quite different situation, see Fig.~\ref{fig:allxF-01}. In this configuration the maximized SSAs within the CGI-GPM approach are strongly suppressed. For the SYY13 LDME set, the situation looks only slightly different since the absence of the CS contribution prevents a further relative cancellation. On the other hand, even maximizing the GSF, the estimates also in this case are already very close to the data. In this respect, within the CGI-GPM approach data at negative $x_F$ appear not very useful to constrain the GSF. This is in contrast to what happens in the GPM framework where, both in CSM and in NRQCD, only a strongly suppressed GSF w.r.t.~its positivity bound could give estimates compatible with PHENIX data.

\begin{figure}[H]
\begin{center}
\includegraphics[trim = 1.cm 0cm 1cm 0cm, width=8.5cm]{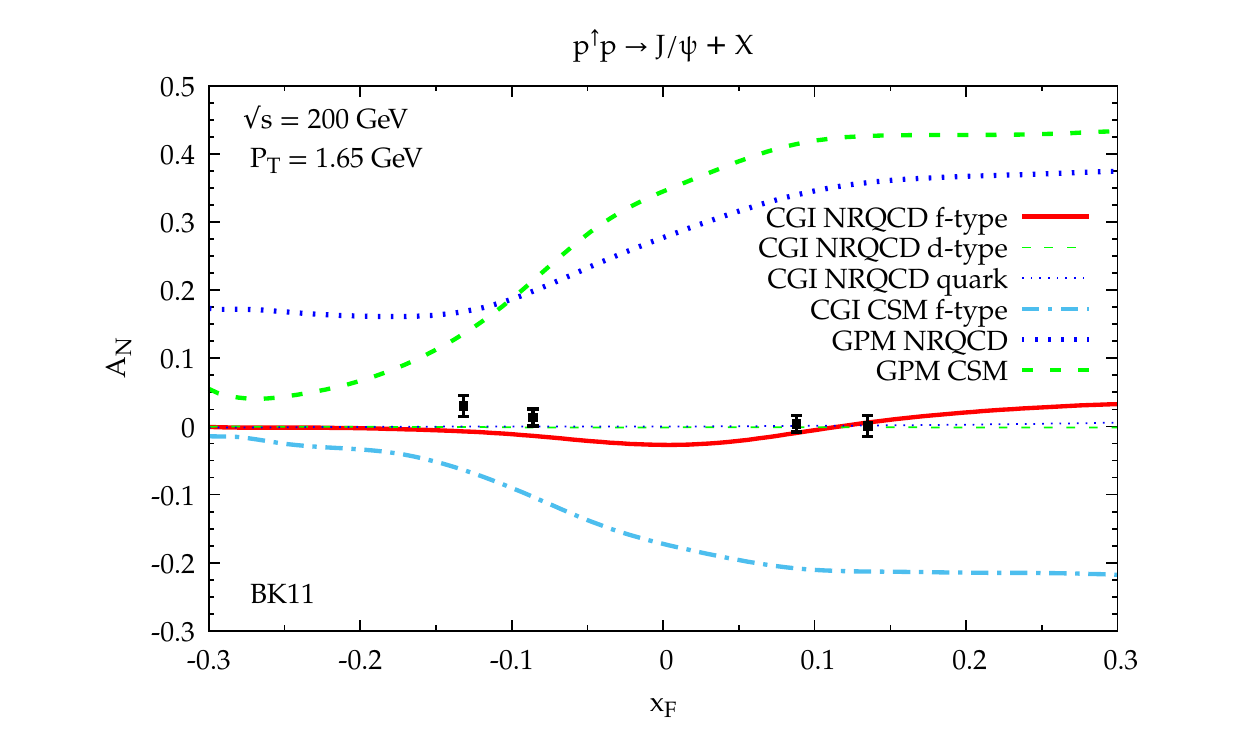}
\includegraphics[trim = 1.cm 0cm 1cm 0cm, width=8.5cm]{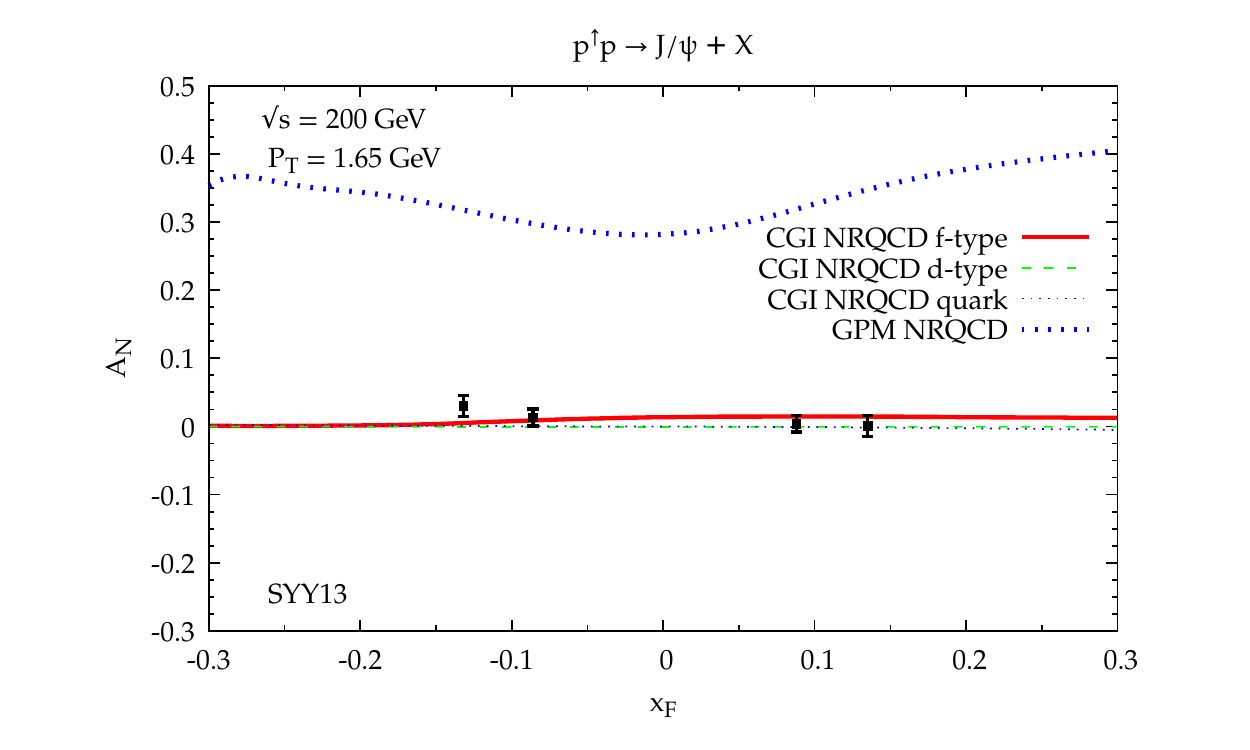}
\end{center}
\caption{Same as in Fig.~\ref{fig:allxF01}  but at fixed $P_T= 1.65$  GeV as a function of $x_F$.}
\label{fig:allpt165}
\end{figure}

Complementary information can be also obtained by looking at the same quantities at fixed $P_T$ (chosen here to be 1.65 GeV as in PHENIX data) as a function of $x_F$. This is shown in Fig.~\ref{fig:allpt165}, where, adopting the same choices as in Figs.~\ref{fig:allxF01} and \ref{fig:allxF-01}, we see that for such $P_T$ values one cannot put any constraint on the $f$-type GSF. A better configuration, in this respect, would be exploring larger $P_T$ values (around 2-3 GeV) and, with some care, very low-$P_T$ values (below 1 GeV), in the positive $x_F$ region (see Fig.~\ref{fig:allxF01}). In such cases the maximized $A_N$ would be sizeable enough and any data could help in constraining the GSF within a CGI-GPM approach.

Before concluding this comparison, we have to mention that the use of the available extraction of the $f$-type (and a fortiori the $d$-type) GSF, from midrapidity pion and $D$-meson SSA data~\cite{DAlesio:2018rnv}, would give results 0.05 (0.15) times smaller than the corresponding maximized estimates, that is almost compatible with zero, and in reasonable agreement with $J/\psi$ data.

For its relevance, we also present estimates for the corresponding $A_N$ in $J/\psi$ production for the kinematics reachable at LHC in the fixed target mode with a transversely polarized target (see the AFTER~\cite{Brodsky:2012vg,Hadjidakis:2018ifr} and LHCSpin~\cite{DiNezza:2019ziv,Aidala:2019pit} proposals at CERN). In such a configuration one could probe even larger light-cone momentum fractions in the polarized proton, accessing the gluon TMDs in a very interesting and complementary region.

\begin{figure}[H]
\begin{center}
\includegraphics[trim = 1.cm 0cm 1cm 0cm, width=8.5cm]{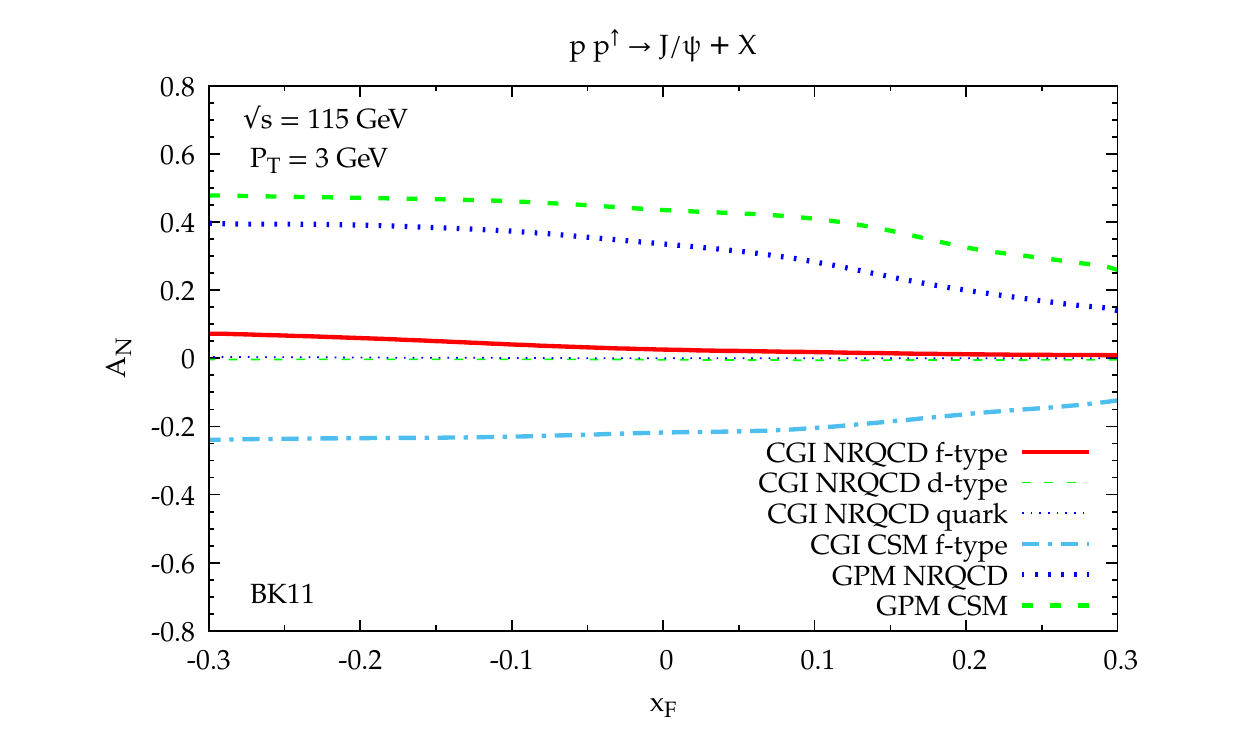}
\includegraphics[trim = 1.cm 0cm 1cm 0cm, width=8.5cm]{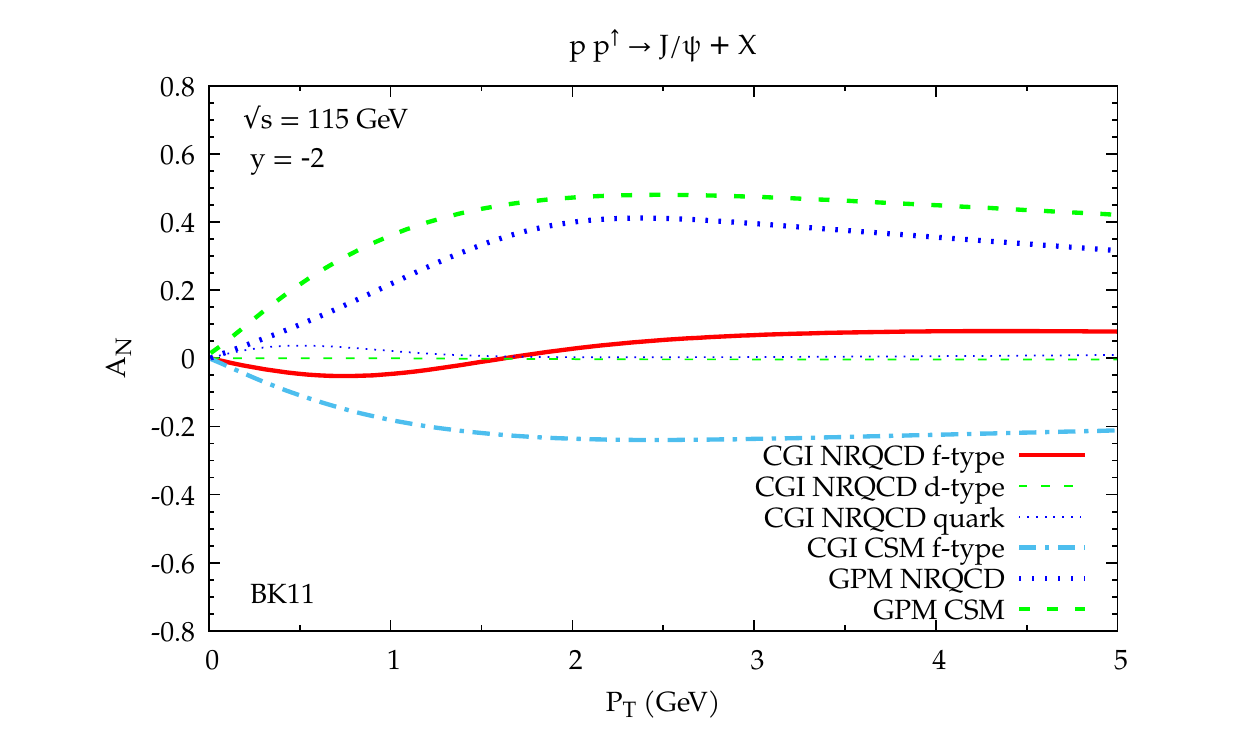}
\end{center}
\caption{Maximized values for $A_N$ for the process $p p^\uparrow \to J/\psi + X$ at $\sqrt s=115$ GeV and $P_T=3$ GeV as a function of $x_F$ (left panel) and at $y=-2$ as a function of $P_T$ (right panel), obtained adopting the CGI-GPM and GPM approaches, within the CS model and NRQCD (BK11 set). Notice that here negative rapidities correspond to the forward region for the polarized proton.}
\label{fig:ANptlhcb}
\end{figure}

In Fig.~\ref{fig:ANptlhcb} we present our maximized estimates for $A_N$ for $p p^\uparrow \to J/\psi + X$ at $\sqrt s=115$ GeV, at fixed $P_T=3$~GeV, as a function of $x_F$ (left panel) and at fixed rapidity $y=-2$, as a function of $P_T$ (right panel), adopting the BK11 set. Notice that in such a configuration the backward rapidity region refers to the forward region for the polarized proton target.
As one can see, at $P_T=3$~GeV (left panel) the maximized contribution from the $f$-type GSF at backward rapidity is around 5\% and, in principle, could be accessed/constrained experimentally. The same is true at very small ($< 1$~GeV) or large ($\ge 3$~GeV) $P_T$ values at $y=-2$ (right panel). We notice that the corresponding estimates, from the $f$-type GSF, at $P_T$ around 2~GeV would be almost negligible at all rapidities. As already discussed in our previous studies, adopting the GPM or the CGI-GPM together with the CSM, the maximized $A_N$ would be much larger and, potentially, easier to constrain.

\section{Conclusions}
\label{sec4}

In this paper we have extended, and somehow completed, a detailed analysis of SSAs for $J/\psi$ production in $pp$ collisions within a phenomenological TMD scheme. This study started in a previous paper, where, employing the Color-Singlet Model for quarkonium formation, we compared the Generalized Parton Model and the Color-Gauge-Invariant GPM. It has been then continued quite recently in a second work, adopting the NRQCD framework within the GPM. Here we have eventually considered its extension within the CGI-GPM. The main interest of this analysis is to see whether and to what extent one can extract information on the poorly known gluon Sivers function, focusing only on this specific process.

We have considered all relevant subprocesses in NRQCD, both for the $2 \to 1$ and the $2 \to 2$ channels, including effects of initial and final state interactions, in the one-gluon-exchange approximation. This leads to the introduction of new color factors, diagram by diagram, and the computation of \emph{modified} hard scattering amplitudes. In such a way one can move the process dependence, coming from ISIs and FSIs, into the hard parts, factorizing the corresponding TMDs. One, well-known, outcome of this approach is the appearance of two independent gluon Sivers functions, referred to as the $d$-type and the $f$-type distributions.

We have then calculated the maximized contributions to $A_N$, separately for the gluon and the quark Sivers effects, adopting the kinematics of the PHENIX experiment, for which data are available.
The main findings are that the quark as well as the $d$-type gluon Sivers functions, even if maximized, give almost negligible contributions to the SSA, leaving at work, as in the CSM, only the $f$-type GSF.
On the other hand, within NRQCD this contribution is also generally quite small and could be relatively sizeable only at forward rapidities and $P_T$ around $2$-$3$ GeV, at least for the two LDME sets considered.

Therefore, while within the GPM, the GSF could be easily constrained by PHENIX SSA data for $J/\psi$ production alone, the situation in the CGI-GPM is quite different. Indeed, if one adopts the CSM, the $f$-type GSF (the only one active) gives still a potentially sizeable contribution; on the contrary, in full NRQCD it could be hardly constrained, and definitely not in the backward region.

We have also presented some maximized estimates of $A_N$, for the kinematics reachable at LHC in a fixed target mode, showing similar features as those discussed for PHENIX setup.

More data, with higher statistics, could certainly help in shedding light on the role of the gluon Sivers function, as well as on its process dependence.

\section*{Acknowledgments}
This work is financially supported by Fondazione Sardegna under the project “Quarkonium at LHC energies”, CUP F71I17000160002 (University of Cagliari). This project has received funding from the European Union's Horizon 2020 research and innovation programme under grant agreement N.~824093.

\appendix

\section{Color factors and amplitudes squared in $p^\uparrow p \to J/\psi + X$ for color-octet states within the CGI-GPM approach}
\label{app:A}

Here we collect all color factors as well as the amplitudes squared for the relevant subprocesses in $p^\uparrow p\to J/\psi + X$ within the CGI-GPM approach (for color-octet states). The color factors and the corresponding amplitudes squared for the color-singlet states can be found in Ref.~\cite{DAlesio:2017rzj}.

The modified amplitudes squared in the CGI-GPM or, more precisely, each contribution to the $\mathcal{M}\mathcal{M}^*$ product between any two of the Feynman diagrams for the specific subprocess, can be written as
\begin{eqnarray}
|\mathcal{M}^{\mathrm{Inc}}|^2= \frac{C^{\mathrm{Inc}}}{C_U}\,|\mathcal{M}^{U}|^2 = \frac{C_I + C_F}{C_U}\,|\mathcal{M}^{U}|^2\,,
\end{eqnarray}
where $\mathcal{M}^{U}$ are the scattering amplitudes for the unpolarized partonic processes.
Here and in what follows, $C_U$ are the color factors when including diagrams entering the unpolarized cross section, as well as the numerator of the SSA in the GPM, (see Fig.~\ref{fig:ppjpsi}a, for the $g+g\to J/\psi + g$ channel), while $C_I$ and $C_F$ are the \emph{new} color factors for ISIs and FSIs respectively (corresponding, still for the $g+g\to J/\psi +g$ channel, to Fig.~\ref{fig:ppjpsi}b and Figs.~\ref{fig:ppjpsi}c,\ref{fig:ppjpsi}d). They can be calculated following the procedure described in Ref.~\cite{DAlesio:2017rzj}. Notice that for subprocesses initiated by a gluon in the polarized proton, the color factors are further distinguished in $C_{I,F}^{(f/d)}$ and, correspondingly, we will have $C^{{\rm Inc}(f/d)} = C_I^{(f/d)} + C_F^{(f/d)} $.

The partonic Mandelstam invariants used below for the $2 \to 2$ channels, $a+b\to J/\psi + c$, and the $2\to 1$ channels, $a+b\to J/\psi$ are so defined:
\be
\hat{s}=(p_a+p_b)^2=2\,p_a\! \cdot\! p_b, ~~~ \hat{t}=(p_a - P_{h})^2=M^2-2\,p_a\! \cdot\! P_{h},~~~
\hat{u}=(p_b-P_{h})^2 = M^2 - 2\,p_b\! \cdot \! P_{h}\,.
\ee

\subsection{$g+g\rightarrow J/\psi +g$ channel}

\begin{figure}[h]
\begin{center}
\includegraphics[height=7cm,width=17cm]{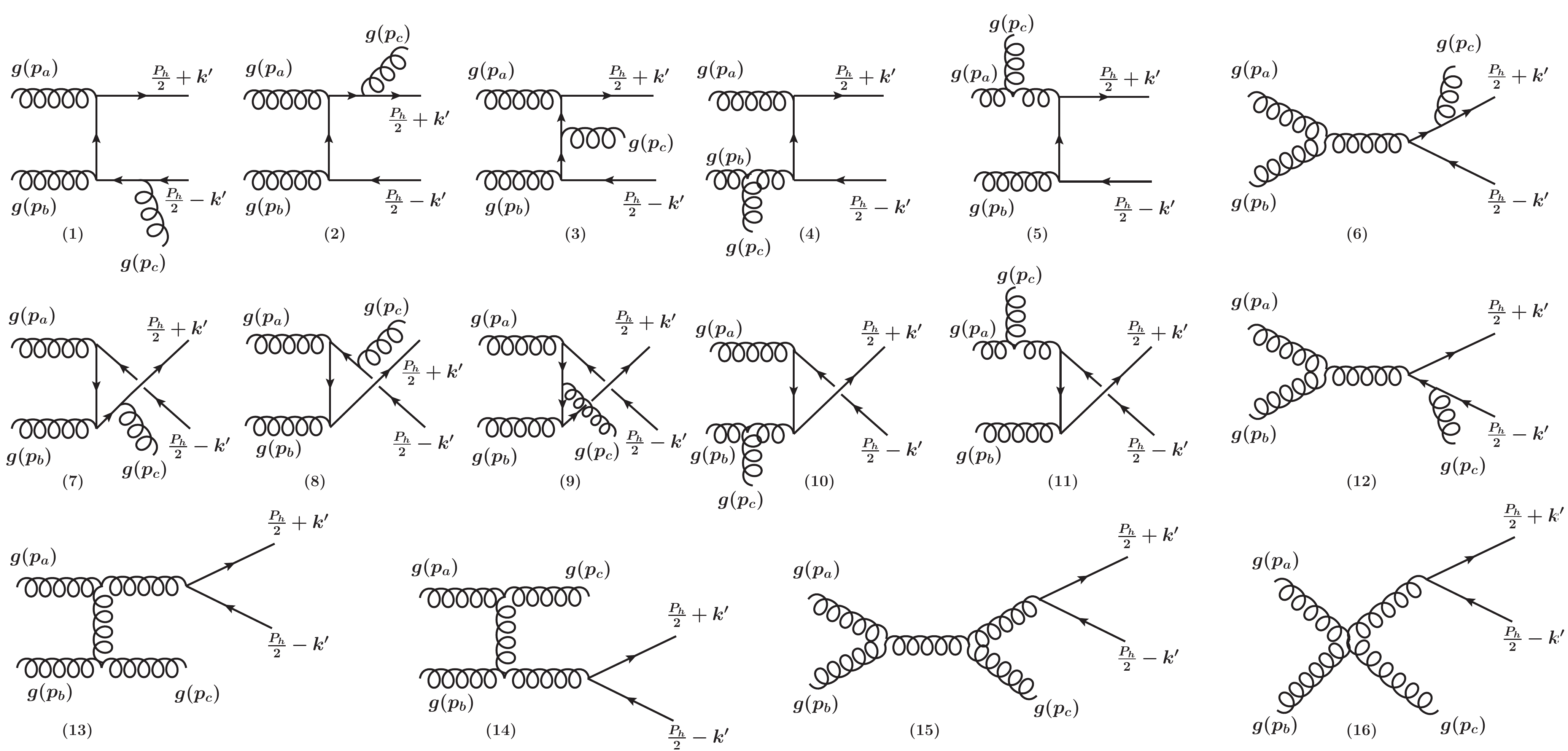}
\end{center}
\caption{Feynman diagrams for the $g+g\rightarrow J/\psi+g$ process.}
\label{fig:gg}
\end{figure}

The set of Feynman diagrams contributing to the $g+g \to J/\psi + g$ channel is shown in Fig.~\ref{fig:gg}. All corresponding color factors are collected in Table~\ref{tab:1} for the $\leftidx{^1}{S}{_0^{(8)}}$ and $\leftidx{^3}{P}{_J^{(8)}}$ states (notice that they are equal) and in Table~\ref{tab:2} for the $\leftidx{^3}{S}{_1^{(8)}}$ state.
The notation in Table~\ref{tab:1} and Table~\ref{tab:2} is the following: in both tables, referring always to Fig.~\ref{fig:gg}, $\circled{A}$ represents the grouping of the $1^{\rm st}$ and the $7^{\rm th}$ Feynman diagrams, $\circled{B}$ of the $2^{\rm nd}$ and the $8^{\rm th}$ and $\circled{C}$ of the $3^{\rm nd}$ and the $9^{\rm th}$. In Table~\ref{tab:1}, $\circled{D}$ implies the grouping of the $4^{\rm th}$ and the $10^{\rm th}$ diagrams, $\circled{E}$ of the $5^{\rm th}$ and the $11^{\rm th}$ and $\circled{F}$ of the $6^{\rm th}$ and the $12^{\rm th}$. Notice that for the $\leftidx{^1}{S}{_0^{(8)}}$ and $\leftidx{^3}{P}{_J^{(8)}}$ states the diagrams from the $13^{\rm th}$ to the $16^{\rm th}$ do not contribute.
In Table~\ref{tab:2}, $\circled{D}$ implies the grouping of the $4^{\rm th}$, the $10^{\rm th}$, the $13^{\rm th}$ and the $16^{\rm th}$ diagrams, $\circled{E}$ of the $5^{\rm th}$, the $11^{\rm th}$, the $14^{\rm th}$ and the $16^{\rm th}$ and $\circled{F}$ of the $6^{\rm th}$, the $12^{\rm th}$, the $15^{\rm th}$ and the $16^{\rm th}$, respectively. As a matter of fact, the $16^{\rm th}$ diagram, which contains the four-gluon vertex, can be split into three parts and then grouped with the diagrams based on the same color factor ($\circled{D}$, $\circled{E}$ and $\circled{F}$).
By the symbol ``$\times$'' we mean, here and in the following, the product of the corresponding amplitudes. All products obtained by crossing, not shown, give the same result. Moreover, all products of group of diagrams for which $C_U=C_I=C_F=0$  are omitted (the same is true for all the following tables).
Finally, the second column in Table~\ref{tab:1} and Table~\ref{tab:2} gives the unpolarized
color factor, $C_U$, while the color factors $C_I^{(f)}$, $C_F^{(f)}$ and $C^{{\rm Inc}(f)}$ are given in the third, fourth and fifth columns respectively. The color factors $C^{(d)}$, not shown, are identically zero.

%
 \begin{table}[t]
  \centering
  \caption{
  Color factors corresponding to the ${^1}{S}{_0^{(8)}}$ and ${^3}{P}{_J^{(8)}}$ states for $g+g\rightarrow J/\psi+g$ channel. See text for further details.
 }
  \label{tab:1}
   \begin{tabular}{ccccc}
   \toprule
  \hline
  \hline
  Diagram &$C_U$ &$C_I^{(f)}$ &$C_F^{(f)}$ &$C^{\mathrm{ Inc}(f)}$\\
 \hline
\midrule
\circled{A}$\times$\circled{A}&$\frac{N^2-4}{4(N^2-1)}$ & $-\frac{N^2-4}{8(N^2-1)}$ & $\frac{N^2-4}{8(N^2-1)}$&0  \\
 \circled{\small{A}}$\times$\circled{D}&$\frac{N^2-4}{4(N^2-1)}$ &$-\frac{N^2-4}{8(N^2-1)}$  &$\frac{N^2-4}{8(N^2-1)}$&0\\
  \circled{A}$\times$\circled{F}&$\frac{N^2-4}{4(N^2-1)}$ & $-\frac{N^2-4}{8(N^2-1)}$  &$\frac{N^2-4}{8(N^2-1)}$ &0\\
  \vspace{0.01cm}\\
     \circled{B}$\times$\circled{B}&$\frac{N^2-4}{4(N^2-1)}$ & $-\frac{N^2-4}{8(N^2-1)}$  & 0 & $-\frac{N^2-4}{8(N^2-1)}$   \\
      \circled{B}$\times$\circled{E}&$\frac{N^2-4}{4(N^2-1)}$ & $-\frac{N^2-4}{8(N^2-1)}$  & 0 & $-\frac{N^2-4}{8(N^2-1)}$  \\
       \circled{B}$\times$\circled{F}&$\frac{N^2-4}{4(N^2-1)}$ & $-\frac{N^2-4}{8(N^2-1)}$  & 0 & $-\frac{N^2-4}{8(N^2-1)}$  \\
 \vspace{0.01cm}\\
   \circled{C}$\times$\circled{C}&$\frac{N^2-4}{4(N^2-1)}$ & 0& $\frac{N^2-4}{8(N^2-1)}$ & $\frac{N^2-4}{8(N^2-1)}$  \\
    \circled{C}$\times$\circled{D}&$-\frac{N^2-4}{4(N^2-1)}$ & 0 & $-\frac{N^2-4}{8(N^2-1)}$ & $-\frac{N^2-4}{8(N^2-1)}$  \\
    \circled{C}$\times$\circled{E}&$-\frac{N^2-4}{4(N^2-1)}$ & 0 & $-\frac{N^2-4}{8(N^2-1)}$ & $-\frac{N^2-4}{8(N^2-1)}$  \\
 \vspace{0.01cm}\\
  \circled{D}$\times$\circled{D}&$\frac{N^2-4}{2(N^2-1)}$ & $-\frac{N^2-4}{8(N^2-1)}$  & $\frac{N^2-4}{4(N^2-1)}$ & $\frac{N^2-4}{8(N^2-1)}$\\
  \circled{D}$\times$\circled{E}&$\frac{N^2-4}{4(N^2-1)}$ &0 & $\frac{N^2-4}{8(N^2-1)}$& $\frac{N^2-4}{8(N^2-1)}$  \\
  \circled{D}$\times$\circled{F}&$\frac{N^2-4}{4(N^2-1)}$ & $-\frac{N^2-4}{8(N^2-1)}$  & $\frac{N^2-4}{8(N^2-1)}$&0  \\
  \vspace{0.01cm}\\
  \circled{E}$\times$\circled{E}&$\frac{N^2-4}{2(N^2-1)}$ & $-\frac{N^2-4}{8(N^2-1)}$  & $\frac{N^2-4}{8(N^2-1)}$ &0  \\
   \circled{E}$\times$\circled{F}&$\frac{N^2-4}{4(N^2-1)}$ & $-\frac{N^2-4}{8(N^2-1)}$  & 0 & $-\frac{N^2-4}{8(N^2-1)}$  \\
   \vspace{0.01cm}\\
      \circled{F}$\times$\circled{F}&$\frac{N^2-4}{2(N^2-1)}$ & $-\frac{N^2-4}{4(N^2-1)}$  & $\frac{N^2-4}{8(N^2-1)}$& $-\frac{N^2-4}{8(N^2-1)}$  \\
  \hline
  \hline
\bottomrule
  \end{tabular}
\end{table}

 \begin{table}[h!]
  \centering
  \caption{Color factors corresponding to the $^{3}{S}{_1^{(8)}}$  state for the $g+g\rightarrow J/\psi+g$ channel. See text for further details.}
  \label{tab:2}
   \begin{tabular}{ccccc}
   \toprule
  \hline
  \hline
  Diagram & $C_U$ &$C_I^{(f)}$ & $C_F^{(f)}$ & $C^{\mathrm{ Inc}(f)}$\\
 \hline
\midrule
\circled{A}$\times$\circled{A}&$\frac{N^4-2 N^2+6}{4 N^2 \left(N^2-1\right)}$ & $-\frac{N^4+4}{8 N^2\left(N^2-1\right)}$  & $\frac{N^4+4}{8 N^2 \left(N^2-1\right)}$ &0 \\
 \circled{A}$\times$\circled{B}&$-\frac{N^2-3}{2 N^2 \left(N^2-1\right)}$ & $-\frac{1}{2 N^2\left( N^2-1\right)}$  & $-\frac{N^2-2}{4 N^2 \left(N^2-1\right)}$&$-\frac{1}{4\left( N^2-1\right)}$\\
\circled{A}$\times$\circled{C}&$-\frac{N^2-3}{2 N^2 \left(N^2-1\right)}$ & $\frac{N^2-2}{4 N^2 \left(N^2-1\right)}$ &$\frac{1}{2 N^2 \left(N^2-1\right)}$&$\frac{1}{4 \left(N^2-1\right)}$\\
 \circled{\small{A}}$\times$\circled{D}&$\frac{N^2}{4 \left(N^2-1\right)}$  &$-\frac{N^2+2}{8\left( N^2-1\right)}$ &$\frac{N^2}{8 \left(N^2-1\right)}$&$-\frac{1}{4\left( N^2-1\right)}$\\
  \circled{A}$\times$\circled{F}&$-\frac{N^2}{4(N^2-1)}$ & $\frac{N^2}{8 \left(N^2-1\right)}$  &$-\frac{N^2+2}{8\left(N^2-1\right)}$&$-\frac{1}{4\left( N^2-1\right)}$ \\
  \vspace{0.01cm}\\
   \circled{B}$\times$\circled{B}&$\frac{N^4-2 N^2+6}{4 N^2 \left(N^2-1\right)}$ &$-\frac{N^4+4}{8 N^2\left( N^2-1\right)}$  &$-\frac{1}{2 N^2}$&$-\frac{N^2+4}{8 \left(N^2-1\right)}$ \\
    \circled{B}$\times$\circled{C}&$-\frac{N^2-3}{2 N^2 \left(N^2-1\right)}$ & $\frac{N^2-2}{4 N^2 \left(N^2-1\right)}$  &$-\frac{N^2-2}{4 N^2 \left(N^2-1\right)}$&0\\
      \circled{B}$\times$\circled{E}&$\frac{N^2}{4 \left(N^2-1\right)}$ &$-\frac{N^2+2}{8\left( N^2-1\right)}$  & $-\frac{1}{4\left(N^2-1\right)}$ &$-\frac{N^2+4}{8 \left(N^2-1\right)}$  \\
       \circled{B}$\times$\circled{F}&$\frac{N^2}{4 \left(N^2-1\right)}$ & $-\frac{N^2}{8\left( N^2-1\right)}$  & $-\frac{1}{4\left(N^2-1\right)}$&$-\frac{N^2+2}{8\left( N^2-1\right)}$ \\
 \vspace{0.01cm}\\
   \circled{C}$\times$\circled{C}&$\frac{N^4-2 N^2+6}{4 N^2 \left(N^2-1\right)}$ & $\frac{1}{2 N^2}$ & $\frac{N^4+4}{8 N^2 \left(N^2-1\right)}$ &$\frac{N^2+4}{8 \left(N^2-1\right)}$ \\
    \circled{C}$\times$\circled{D}&$-\frac{N^2}{4\left( N^2-1\right)}$ & $-\frac{1}{4\left( N^2-1\right)}$ & $-\frac{N^2}{8\left( N^2-1\right)}$&$-\frac{N^2+2}{8\left( N^2-1\right)}$  \\
    \circled{C}$\times$\circled{E}&$-\frac{N^2}{4\left( N^2-1\right)}$ & $-\frac{1}{4\left( N^2-1\right)}$ & $-\frac{N^2+2}{8\left( N^2-1\right)}$&$-\frac{N^2+4}{8\left( N^2-1\right)}$  \\
 \vspace{0.01cm}\\
  \circled{D}$\times$\circled{D}&$\frac{N^2}{2(N^2-1)}$ & $-\frac{N^2}{8(N^2-1)}$  & $\frac{N^2}{4(N^2-1)}$&$\frac{N^2}{8 (N^2-1)}$  \\
  \circled{D}$\times$\circled{E}&$\frac{N^2}{4(N^2-1)}$ &0 & $\frac{N^2}{8(N^2-1)}$ & $\frac{N^2}{8(N^2-1)}$ \\
  \circled{D}$\times$\circled{F}&$-\frac{N^2}{4(N^2-1)}$ & $\frac{N^2}{8(N^2-1)}$  & $-\frac{N^2}{8(N^2-1)}$ &0 \\
  \vspace{0.01cm}\\
  \circled{E}$\times$\circled{E}&$\frac{N^2}{2(N^2-1)}$& $-\frac{N^2}{8(N^2-1)}$  & $\frac{N^2}{8(N^2-1)}$ &0  \\
   \circled{E}$\times$\circled{F}&$\frac{N^2}{4(N^2-1)}$ & $-\frac{N^2}{8(N^2-1)}$  & 0 & $-\frac{N^2}{8(N^2-1)}$  \\
   \vspace{0.01cm}\\
      \circled{F}$\times$\circled{F}&$\frac{N^2}{2(N^2-1)}$ & $-\frac{N^2}{4(N^2-1)}$  & $\frac{N^2}{8(N^2-1)}$ & $-\frac{N^2}{8(N^2-1)}$ \\
  \hline
  \hline
\bottomrule
  \end{tabular}
\end{table}


Collecting together all contributions of Fig.~\ref{fig:gg}, with the appropriate color factors from Tables \ref{tab:1}-\ref{tab:2}, we find the following expressions for the amplitudes squared:
\be
 |\mathcal{M}^{\mathrm{Inc}(f)}[\leftidx{^{1}}{S}{_0}^{(8)}]|^2&=&
\frac{(4\pi \alpha_s)^3\langle 0\mid \mathcal{O}_8^{J/\psi}(\leftidx{^1}{S}{_0})\mid 0\rangle}
{64 \hat{s} \hat{t} (\hat{s}+\hat{u})^2 (\hat{s}+\hat{t})^2 (\hat{u}+\hat{t})^2 M}
\big\{5 (\hat{s}-\hat{t}) (\hat{s} (\hat{u}+2 \hat{t})+\hat{u} \hat{t}) \big(\hat{s}^4\nonumber\\
&&{}+2 \hat{s}^3 (\hat{u}+\hat{t})+3 \hat{s}^2 (\hat{u}+\hat{t})^2+2 \hat{s} (\hat{u}+\hat{t})^3+\big(\hat{u}^2+\hat{u}
   \hat{t}+\hat{t}^2\big)^2\big)\big\}\,,
 \ee

 \be
 |\mathcal{M}^{\mathrm{Inc}(f)}[\leftidx{^{3}}{S}{_1}^{(8)}]|^2&=&
 -\frac{(4\pi \alpha_s)^3\langle 0\mid \mathcal{O}_8^{J/\psi}(\leftidx{^3}{S}{_1})\mid 0\rangle}
 {96 (\hat{s}+\hat{u})^2 (\hat{s}+\hat{t})^2 (\hat{u}+\hat{t})^2 M^3}
 \big\{(\hat{s}-\hat{t}) \big(17 \hat{s}^3 \big(\hat{t}^2+\hat{u} \hat{t}+\hat{u}^2\big)+\hat{s}^2 \big(26 \hat{u}^3\nonumber\\
&&{}+66 \hat{u}^2 \hat{t}+54 \hat{u} \hat{t}^2+17 \hat{t}^3\big)+\hat{s} \hat{u} \hat{t} \big(38 \hat{u}^2+66 \hat{u} \hat{t}+17
   \hat{t}^2\big)+\hat{u}^2 \hat{t}^2 (26 \hat{u}+17 \hat{t})\big)
 \big\}\,,
  \ee

\be
 |\mathcal{M}^{\mathrm{Inc}(f)}[\leftidx{^{3}}{P}{_0}^{(8)}]|^2&=&
\frac{(4\pi \alpha_s)^3\langle 0\mid \mathcal{O}_8^{J/\psi}(\leftidx{^3}{P}{_0})\mid 0\rangle}
{48\hat{s}\hat{t}(\hat{s}+\hat{u})^4 (\hat{s}+\hat{t})^4 (\hat{u}+\hat{t})^4 M^3}
\big\{5 (\hat{s}-\hat{t}) \big(9 \hat{s}^9 (\hat{u}+\hat{t})^2 (\hat{u}+2 \hat{t})
+3 \hat{s}^8 (\hat{u}+\hat{t})^2 \big(12 \hat{u}^2\nonumber\\
&&{}+39 \hat{u} \hat{t}+22 \hat{t}^2\big)+2
   \hat{s}^7 \big(36 \hat{u}^5+237 \hat{u}^4 \hat{t}+556 \hat{u}^3 \hat{t}^2+607 \hat{u}^2 \hat{t}^3
   +310 \hat{u} \hat{t}^4+60 \hat{t}^5\big)\nonumber\\
   &&{}+2 \hat{s}^6 \big(45   \hat{u}^6+369 \hat{u}^5 \hat{t}+1108 \hat{u}^4 \hat{t}^2+1638 \hat{u}^3 \hat{t}^3+1257 \hat{u}^2 \hat{t}^4
   +477 \hat{u} \hat{t}^5+72 \hat{t}^6\big)\nonumber\\
   &&{}+2 \hat{s}^5 \big(36\hat{u}^7+366 \hat{u}^6 \hat{t}+1384 \hat{u}^5 \hat{t}^2+2658 \hat{u}^4 \hat{t}^3+2795 \hat{u}^3 \hat{t}^4+1604 \hat{u}^2 \hat{t}^5+477 \hat{u} \hat{t}^6+60
   \hat{t}^7\big)\nonumber\\
   &&{}+\hat{s}^4 \big(36 \hat{u}^8+456 \hat{u}^7 \hat{t}+2140 \hat{u}^6 \hat{t}^2+5174 \hat{u}^5 \hat{t}^3+7092 \hat{u}^4 \hat{t}^4
   +5590 \hat{u}^3
   \hat{t}^5+2514 \hat{u}^2 \hat{t}^6\nonumber\\
   &&{}+620 \hat{u} \hat{t}^7+66 \hat{t}^8\big)+\hat{s}^3 \big(9 \hat{u}^9+168 \hat{u}^8 \hat{t}+1008 \hat{u}^7 \hat{t}^2+3004 \hat{u}^6
   \hat{t}^3+5174 \hat{u}^5 \hat{t}^4+5316 \hat{u}^4 \hat{t}^5\nonumber\\
   &&{}+3276 \hat{u}^3 \hat{t}^6
   +1214 \hat{u}^2 \hat{t}^7+249 \hat{u} \hat{t}^8+18 \hat{t}^9\big)+\hat{s}^2 \hat{u} \hat{t}
   \big(27 \hat{u}^8+264 \hat{u}^7 \hat{t}+1008 \hat{u}^6 \hat{t}^2\nonumber\\
   &&{}+2140 \hat{u}^5 \hat{t}^3
   +2768 \hat{u}^4 \hat{t}^4+2216 \hat{u}^3 \hat{t}^5+1112 \hat{u}^2 \hat{t}^6+336   \hat{u} \hat{t}^7+45 \hat{t}^8\big)+3 \hat{s}
   \hat{u}^2 \hat{t}^2 (\hat{u}\nonumber\\
&&{}   +\hat{t}) \big(9 \hat{u}^6+47 \hat{u}^5 \hat{t}+105 \hat{u}^4 \hat{t}^2+139 \hat{u}^3 \hat{t}^3
   +107 \hat{u}^2   \hat{t}^4+51 \hat{u} \hat{t}^5+12 \hat{t}^6\big)\nonumber\\
&&{}   +9 \hat{u}^3 \hat{t}^3 (\hat{u}+\hat{t})^2 \big(\hat{u}^2+\hat{u} \hat{t}+\hat{t}^2\big)^2\big)
\big\}\,,
   \ee

 \be
 |\mathcal{M}^{\mathrm{Inc}(f)}[\leftidx{^{3}}{P}{_1}^{(8)}]|^2&=&
 \frac{5(4\pi \alpha_s)^3\langle 0\mid \mathcal{O}_8^{J/\psi}(\leftidx{^3}{P}{_1})\mid 0\rangle}
{24 (\hat{s}+\hat{u})^4 (\hat{s}+\hat{t})^4   (\hat{u}+\hat{t})^4 M^3}
\big\{2 \hat{s}^4 \hat{u}^6 (\hat{s}+\hat{u})+\hat{s} \hat{t}^7 (2 \hat{s}+\hat{u}) \big(\hat{s}^2+\hat{s} \hat{u}+\hat{u}^2\big)-\hat{s}^3
   \hat{u}^2 \hat{t}^2 \big(3 \hat{s}^2\nonumber\\
    &&{}+3 \hat{s} \hat{u}-2 \hat{u}^2\big) \big(\hat{s}^2+4 \hat{s} \hat{u}+2 \hat{u}^2\big)+\hat{s}
   \hat{t}^6 \big(2 \hat{s}^4+12 \hat{s}^3 \hat{u}+19 \hat{s}^2 \hat{u}^2+15 \hat{s} \hat{u}^3+4 \hat{u}^4\big)+\hat{s}^3 \hat{u}^3 \hat{t}
   \big(-\hat{s}^4\nonumber\\
    &&{}-4 \hat{s}^3 \hat{u}+2 \hat{s}^2 \hat{u}^2+8 \hat{s} \hat{u}^3+2 \hat{u}^4\big)-\hat{s} \hat{u} \hat{t}^3 \big(3 \hat{s}^6+19
   \hat{s}^5 \hat{u}+33 \hat{s}^4 \hat{u}^2+16 \hat{s}^3 \hat{u}^3+4 \hat{s} \hat{u}^5+2 \hat{u}^6\big)\nonumber\\
    &&{}+\hat{t}^5 \big(-2 \hat{s}^6+18 \hat{s}^4
   \hat{u}^2+33 \hat{s}^3 \hat{u}^3+16 \hat{s}^2 \hat{u}^4-2 \hat{s} \hat{u}^5-2 \hat{u}^6\big)-2 \hat{t}^4 \big(\hat{s}^7+6 \hat{s}^6 \hat{u}+9
   \hat{s}^5 \hat{u}^2-8 \hat{s}^3 \hat{u}^4\nonumber\\
    &&{}+\hat{s}^2 \hat{u}^5+4 \hat{s} \hat{u}^6+\hat{u}^7\big)\big\}\,,
   \ee

\be
 |\mathcal{M}^{\mathrm{Inc}(f)}[\leftidx{^{3}}{P}{_2}^{(8)}]|^2&=&
 \frac{(4\pi \alpha_s)^3\langle 0\mid \mathcal{O}_8^{J/\psi}(\leftidx{^3}{P}{_2})\mid 0\rangle}
 {24 \hat{s} \hat{t}
   (\hat{s}+\hat{u})^4 (\hat{s}+\hat{t})^4 (\hat{u}+\hat{t})^4 M^3}
   \big\{(\hat{s}-\hat{t}) \big(6 \hat{s}^9 (\hat{u}+\hat{t})^2 (\hat{u}+2 \hat{t})
   +6 \hat{s}^8 (\hat{u}+\hat{t}) \big(4 \hat{u}^3\nonumber\\
    &&{}+17 \hat{u}^2 \hat{t}+20 \hat{u} \hat{t}^2+9
   \hat{t}^3\big)+\hat{s}^7 \big(48 \hat{u}^5+288 \hat{u}^4 \hat{t}+617 \hat{u}^3 \hat{t}^2
   +695 \hat{u}^2 \hat{t}^3+431 \hat{u} \hat{t}^4+114 \hat{t}^5\big)\nonumber\\
    &&{}+\hat{s}^6   \big(60 \hat{u}^6+402 \hat{u}^5 \hat{t}+968 \hat{u}^4 \hat{t}^2+1326 \hat{u}^3 \hat{t}^3+1194 \hat{u}^2 \hat{t}^4+639 \hat{u} \hat{t}^5+144 \hat{t}^6\big)+\hat{s}^5
   \big(48 \hat{u}^7\nonumber\\
    &&{}+384 \hat{u}^6 \hat{t}+1052 \hat{u}^5 \hat{t}^2+1602 \hat{u}^4 \hat{t}^3+1759 \hat{u}^3 \hat{t}^4
   +1396 \hat{u}^2 \hat{t}^5+639 \hat{u} \hat{t}^6+114
   \hat{t}^7\big)+\hat{s}^4 \big(24 \hat{u}^8\nonumber\\
    &&{}+252 \hat{u}^7 \hat{t}+856 \hat{u}^6 \hat{t}^2+1484 \hat{u}^5 \hat{t}^3+1800 \hat{u}^4 \hat{t}^4
   +1759 \hat{u}^3   \hat{t}^5+1194 \hat{u}^2 \hat{t}^6+431 \hat{u} \hat{t}^7+54 \hat{t}^8\big)\nonumber\\
    &&{}+\hat{s}^3 \big(6 \hat{u}^9+102 \hat{u}^8 \hat{t}+486 \hat{u}^7 \hat{t}^2+1066 \hat{u}^6
   \hat{t}^3+1484 \hat{u}^5 \hat{t}^4+1602 \hat{u}^4 \hat{t}^5+1326 \hat{u}^3 \hat{t}^6
   +695 \hat{u}^2 \hat{t}^7\nonumber\\
    &&{}+174 \hat{u} \hat{t}^8+12 \hat{t}^9\big)+\hat{s}^2 \hat{u} \hat{t}
   \big(18 \hat{u}^8+156 \hat{u}^7 \hat{t}+486 \hat{u}^6 \hat{t}^2+856 \hat{u}^5 \hat{t}^3
   +1052 \hat{u}^4 \hat{t}^4+968 \hat{u}^3 \hat{t}^5\nonumber\\
    &&{}+617 \hat{u}^2 \hat{t}^6+222 \hat{u}
   \hat{t}^7+30 \hat{t}^8\big)+6 \hat{s} \hat{u}^2 \hat{t}^2 (\hat{u}+\hat{t}) \big(3 \hat{u}^6
   +14 \hat{u}^5 \hat{t}+28 \hat{u}^4 \hat{t}^2+36 \hat{u}^3 \hat{t}^3+31 \hat{u}^2
   \hat{t}^4\nonumber\\
    &&{}+17 \hat{u} \hat{t}^5+4 \hat{t}^6\big)+6 \hat{u}^3 \hat{t}^3 (\hat{u}+\hat{t})^2
   \big(\hat{u}^2+\hat{u} \hat{t}+\hat{t}^2\big)^2\big)
   \big\}\,.
   \ee
Assuming the validity of the common heavy-quark spin symmetry relations~\cite{Bodwin:1994jh}
\begin{align}
\label{eq:hqss}
\langle 0 \vert{\cal  O}_8^{J/\psi} (^3P_J)\vert  0 \rangle = (2J+1) \,\langle 0 \vert {\cal O}_8^{J/\psi} (^3P_0)\vert 0 \rangle\,,
\end{align}
the sum of the last three contributions simplifies a lot:
\be
|\mathcal{M}^{\mathrm{Inc}(f)}[\leftidx{^{3}}{P}_{[J]}^{(8)}]|^2&=&
\frac{(4\pi \alpha_s)^3\langle 0\mid \mathcal{O}_8^{J/\psi}(\leftidx{^3}{P}{_0})\mid 0\rangle}
{16\hat{s}\hat{t}(\hat{s}+\hat{u})^3 (\hat{s}+\hat{t})^3 (\hat{u}+\hat{t})^3 M^3}
\big\{5 (\hat{s}-\hat{t}) \big(7 \hat{s}^7 (\hat{u}+\hat{t}) (\hat{u}+2 \hat{t})+\hat{s}^6 \big(21 \hat{u}^3+91 \hat{u}^2 \hat{t}\nonumber\\
&&{}+106 \hat{u} \hat{t}^2+44 \hat{u}^3\big)+\hat{s}^5
   \big(35 \hat{u}^4+175 \hat{u}^3 \hat{t}+268 \hat{u}^2 \hat{t}^2+212 \hat{u} \hat{t}^3+68 \hat{t}^4\big)+\hat{s}^4 \big(35 \hat{u}^5\nonumber\\
&&{}+213 \hat{u}^4 \hat{t}+398
   \hat{u}^3 \hat{t}^2+426 \hat{u}^2 \hat{t}^3+254 \hat{u} \hat{t}^4+68 \hat{t}^5\big)+\hat{s}^3 \big(21 \hat{u}^6+165 \hat{u}^5 \hat{t}+394 \hat{u}^4 \hat{t}^2\nonumber\\
&&{}+528 \hat{u}^3
   \hat{t}^3+426 \hat{u}^2 \hat{t}^4+212 \hat{u} \hat{t}^5+44 \hat{t}^6\big)+\hat{s}^2 \big(7 \hat{u}^7+75 \hat{u}^6 \hat{t}+236 \hat{u}^5 \hat{t}^2+394 \hat{u}^4
   \hat{t}^3\nonumber\\
&&{}+398 \hat{u}^3 \hat{t}^4+268 \hat{u}^2 \hat{t}^5+106 \hat{u} \hat{t}^6+14 \hat{t}^7\big)+\hat{s} \hat{u} \hat{t} \big(14 \hat{u}^6+75 \hat{u}^5 \hat{t}+165 \hat{u}^4
   \hat{t}^2+213 \hat{u}^3 \hat{t}^3\nonumber\\
&&{}+175 \hat{u}^2 \hat{t}^4+91 \hat{u} \hat{t}^5+21 \hat{t}^6\big)+7 \hat{u}^2 \hat{t}^2 (\hat{u}+\hat{t}) \big(\hat{u}^2+\hat{u}\hat{t}+\hat{t}^2\big)^2\big)
\big\}\,,
   \ee
where the symbol $[J]$ means a sum over $J=0,1,2$.
\subsection{$g+q\rightarrow J/\psi +q$ channel}
\begin{figure}[h]
\begin{center}
\includegraphics[height=6cm,width=15cm]{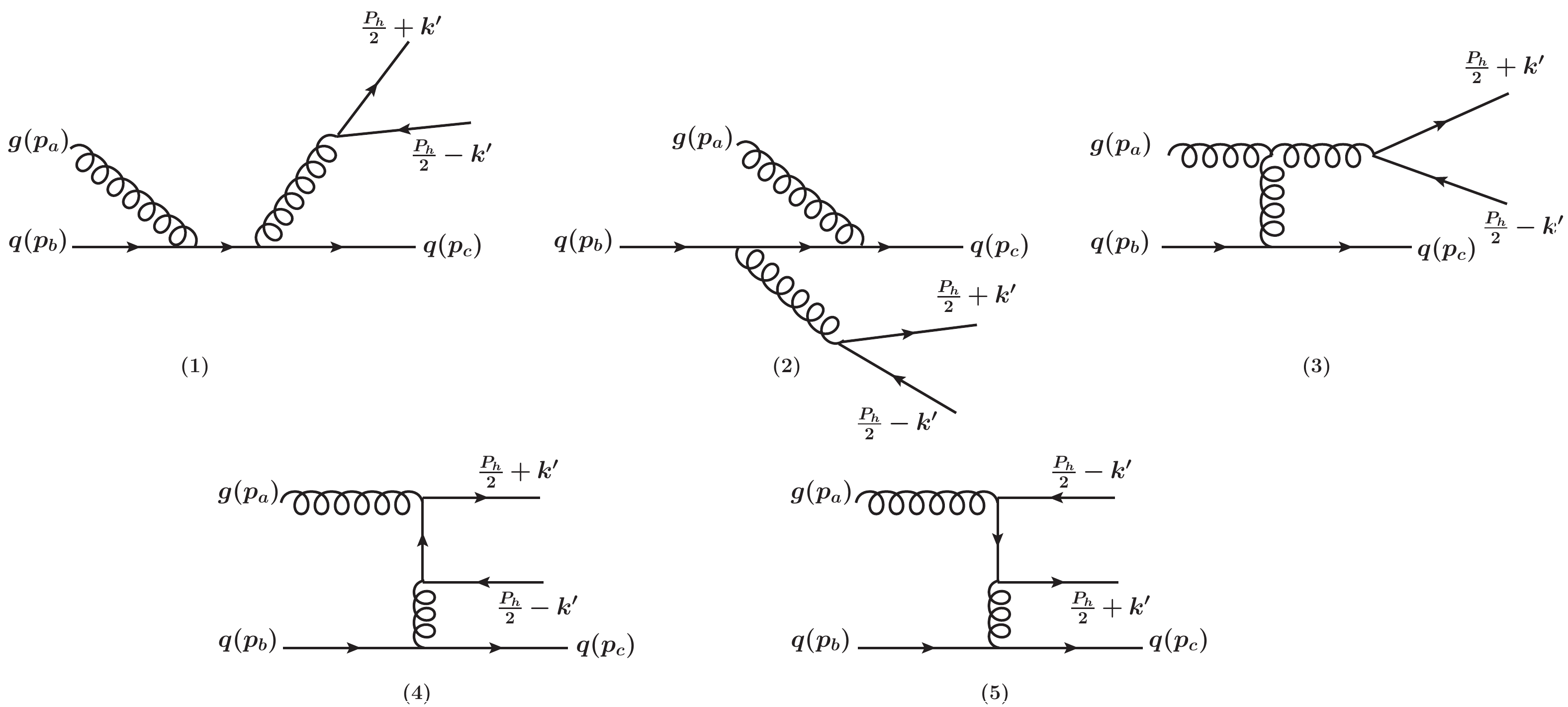}
\end{center}
\caption{\label{fig_gq}Feynman diagrams for the $g+q\rightarrow J/\psi+q$ process.}
\end{figure}

The Feynman diagrams contributing to the $g+q \to J/\psi + q$ channel are shown in Fig.~\ref{fig_gq}, while the corresponding color factors are collected in Table~\ref{table_gq} (upper signs).

Here, as well as in the following Table~\ref{table_qg}, the label \circled{A} refers to the $1^{\rm st}$ Feynman diagram, \circled{B} to the $2^{\rm nd}$, while \circled{C} represents the grouping of the $3^{\rm rd}$, $4^{\rm th}$ and $5^{\rm th}$ diagrams in Fig.~{\ref{fig_gq}}. For the $\leftidx{^1}{S}{_0^{(8)}}$ and  $\leftidx{^3}{P}{_J^{(8)}}$ states, the $1^{\rm st}$, $2^{\rm nd}$ and $3^{\rm rd}$ Feynman diagrams in Fig.~{\ref{fig_gq}} do not contribute.

 \begin{table}[h]
  \centering
 \caption{Color factors corresponding to the ${^3}{S}{_1^{(8)}}$, ${^1}{S}{_0^{(8)}}$ and ${^3}{P}{_J^{(8)}}$ states for the $g+q\rightarrow J/\psi+q$ (upper signs) and $g+\bar q\rightarrow J/\psi+\bar q$ (lower signs) channels.}
  \label{table_gq}
   \begin{tabular}{ccccccccc}
   \toprule
  \hline
  \hline
  State  &Diagram &$C_U$ &$C_I^{(f)}$ &$C_I^{(d)}$  &$C_F^{(f)}$  &$C_F^{(d)}$  &$C^{\mathrm{ Inc}(f)}$
 &$C^{\mathrm{ Inc}(d)}$\\
 \hline
\midrule
\vspace*{0.04cm}$\leftidx{^3}{S}{_1^{(8)}}$
&\circled{A}$\times$\circled{A}&$\frac{N^2-1}{8 N^2}$ & $-\frac{N^2-1}{16N^2}$ & $\pm\frac{N^2-1}{16N^2}$ & $\frac{1}{16}$ &$\pm\frac{1}{16}$
& $\frac{1}{16 N^2}$ & $\pm\frac{2N^2-1}{16N^2}$\vspace*{0.04cm}\\
 $~~$ &\circled{A}$\times$\circled{B}&$-\frac{1}{8 N^2}$ & $\frac{1}{16 N^2}$ &  $\mp\frac{1}{16 N^2}$ &0&0& $\frac{1}{16 N^2}$&  $\mp\frac{1}{16 N^2}$\vspace*{0.04cm}\\
$~~$ &\circled{A}$\times$\circled{C}&$\pm \frac{1}{8}$ & $\mp\frac{1}{16}$ & $\frac{1}{16}$ &$\pm\frac{1}{16}$&$\frac{1}{16}$&0&$\frac{1}{8}$ \\

  \vspace{0.01cm}\\
  $~~$ & \circled{B}$\times$\circled{B}&$\frac{N^2-1}{8 N^2}$ & $\frac{1}{16 N^2}$ &$\mp\frac{1}{16 N^2}$ & $\frac{1}{16}$  &$\mp\frac{1}{16}$& $\frac{N^2+1}{16 N^2}$& $\mp\frac{N^2+1}{16 N^2}$\vspace*{0.04cm}  \\
  $~~$ &  \circled{B}$\times$\circled{C}&$\mp\frac{1}{8}$ & 0 & 0 &$\mp\frac{1}{16}$&$\frac{1}{16}$&$\mp\frac{1}{16}$&$\frac{1}{16}$\\

 \vspace{0.01cm}\\
   $~~$ &\circled{C}$\times$\circled{C}&$\frac{1}{4}$ & $-\frac{1}{16}$& $\pm\frac{1}{16}$& $\frac{1}{8}$ &0& $\frac{1}{16}$& $\pm\frac{1}{16}$  \\

   \vspace{0.01cm}\\
   $\leftidx{^1}{S}{_0^{(8)}}$\&  $\leftidx{^3}{P}{_J^{(8)}}$& \circled{C}$\times$\circled{C}& $\frac{N^2-4}{4N^2}$ &
  $-\frac{N^2-4}{16N^2}$ &$\pm\frac{N^2-12}{16N^2}$& $\frac{N^2-4}{8N^2}$ &0& $\frac{N^2-4}{16N^2}$&$\pm\frac{N^2-12}{16N^2}$\vspace*{0.04cm}\\
  \hline
  \hline
\bottomrule
  \end{tabular}
\end{table}

Collecting together all contributions of Fig.~\ref{fig_gq} with the appropriate color factors from Table~\ref{table_gq}, we find the following expressions for the amplitudes squared:

\begin{eqnarray}\label{gq1s0f}
|\mathcal{M}^{\mathrm{Inc}(f)}[\leftidx{^{1}}{S}{_0}^{(8)}]|^2=-\frac{5(4\pi\alpha_s)^3}{288M}
\langle 0\mid \mathcal{O}_8^{J/\psi}(\leftidx{^1}{S}{_0})\mid 0\rangle
\frac{\hat{s}^2+\hat{u}^2}{\hat{t}(\hat{t}-M^2)^2}
\,,
\end{eqnarray}
\begin{eqnarray}\label{gq1s0d}
|\mathcal{M}^{\mathrm{Inc}(d)}[\leftidx{^{1}}{S}{_0}^{(8)}]|^2=-\frac{3}{5} |\mathcal{M}^{\mathrm{Inc}(f)}[\leftidx{^{1}}{S}{_0}^{(8)}]|^2\,,
\end{eqnarray}
\begin{equation}\label{gq3s1f}
\begin{split}
|\mathcal{M}^{\mathrm{Inc}(f)}[\leftidx{^{3}}{S}{_1}^{(8)}]|^2=&-\frac{(4\pi\alpha_s)^3 }{432M^3}
\langle 0\mid \mathcal{O}_8^{J/\psi}(\leftidx{^3}{S}{_1})\mid 0\rangle\,
 \frac{\left(10\hat{s}^2+2\hat{s}\hat{u}+\hat{u}^2\right) \left(\hat{s}^2+2 \hat{s} \hat{t}+\hat{u}^2+2 \hat{t} (\hat{u}+\hat{t})\right)}{\hat{ s}\hat{u} (\hat{s}+\hat{u})^2}
\,,
\end{split}
\end{equation}
\begin{equation}\label{gq3s1d}
\begin{split}
|\mathcal{M}^{\mathrm{Inc}(d)}[\leftidx{^{3}}{S}{_1}^{(8)}]|^2=&\frac{(4\pi\alpha_s)^3 }{432M^3}
\langle 0\mid \mathcal{O}_8^{J/\psi}(\leftidx{^3}{S}{_1})\mid 0\rangle\,
 \frac{\left(10\hat{s}^2+2\hat{s}\hat{u}-17\hat{u}^2\right) \left(\hat{s}^2+2 \hat{s} \hat{t}+\hat{u}^2+2 \hat{t} (\hat{u}+\hat{t})\right)}{\hat{ s}\hat{u} (\hat{s}+\hat{u})^2}
\,,
\end{split}
\end{equation}
\begin{eqnarray}\label{gq3p0f}
|\mathcal{M}^{\mathrm{Inc}(f)}[\leftidx{^{3}}{P}{_0}^{(8)}]|^2=-\frac{5(4\pi\alpha_s)^3}{216M^3}
\langle 0\mid \mathcal{O}_8^{J/\psi}(\leftidx{^3}{P}{_0})\mid 0\rangle\,
\frac{(\hat{t}-3M^2)^2(\hat{s}^2+\hat{u}^2)}{\hat{t}(\hat{t}-M^2)^4}
\,,
\end{eqnarray}
\begin{eqnarray}\label{gq3p0d}
|\mathcal{M}^{\mathrm{Inc}(d)}[\leftidx{^{3}}{P}{_0}^{(8)}]|^2=-\frac{3}{5}
|\mathcal{M}^{\mathrm{Inc}(f)}[\leftidx{^{3}}{P}{_0}^{(8)}]|^2\,,
\end{eqnarray}
\begin{eqnarray}\label{gq3p1f}
|\mathcal{M}^{\mathrm{Inc}(f)}[\leftidx{^{3}}{P}{_1}^{(8)}]|^2=-\frac{5(4\pi\alpha_s)^3}{108M^3}
\langle 0\mid \mathcal{O}_8^{J/\psi}(\leftidx{^3}{P}{_1})\mid 0\rangle\,
\frac{\hat{t} \left(\hat{s}^2+4 \hat{s} \hat{u}+\hat{u}^2\right)+4\hat{ s} \hat{u} (\hat{s}+\hat{u})}{ (\hat{s}+\hat{u})^4 }
\,,
\end{eqnarray}
\begin{eqnarray}\label{gq3p1d}
|\mathcal{M}^{\mathrm{Inc}(d)}[\leftidx{^{3}}{P}{_1}^{(8)}]|^2=-\frac{3}{5}
|\mathcal{M}^{\mathrm{Inc}(f)}[\leftidx{^{3}}{P}{_1}^{(8)}]|^2\,,
\end{eqnarray}
\be\label{gq3p2f}
|\mathcal{M}^{\mathrm{Inc}(f)}[\leftidx{^{3}}{P}{_2}^{(8)}]|^2&=&-\frac{(4\pi\alpha_s)^3}{108M^3}
\langle 0\mid \mathcal{O}_8^{J/\psi}(\leftidx{^3}{P}{_2})\mid 0\rangle \nonumber\\
&\times &{}\frac{\hat{t}^2 \left(7 \hat{s}^2+12 \hat{s} \hat{u}+7 \hat{u}^2\right)+12\hat{ u} \left(\hat{s}^2+\hat{s} \hat{u}+\hat{u}^2\right)
(\hat{s}+\hat{u})+6
   \left(\hat{s}^2+\hat{u}^2\right) (\hat{s}+\hat{u})^2}{\hat{t} (\hat{s}+\hat{u})^4}\,,
\ee
\begin{eqnarray}\label{gq3p2d}
|\mathcal{M}^{\mathrm{Inc}(d)}[\leftidx{^{3}}{P}{_2}^{(8)}]|^2=-\frac{3}{5}
|\mathcal{M}^{\mathrm{Inc}(f)}[\leftidx{^{3}}{P}{_2}^{(8)}]|^2\,.
\end{eqnarray}

Once again, by applying the relation in Eq.~(\ref{eq:hqss}), we have
\be\label{gq3pf}
|\mathcal{M}^{\mathrm{Inc}(f)}[\leftidx{^{3}}{P}_{[J]}^{(8)}]|^2& = & \frac{5(4\pi\alpha_s)^3}{72M^3 \hat{t}
(\hat{s}+\hat{u})^3}
\langle 0 \mid \mathcal{O}_8^{J/\psi}(\leftidx{^3}{P}{_0})\mid 0\rangle\,
 \big\{-4 \hat{t} \big(3 \hat{s}^2+4 \hat{s} \hat{u}+3 \hat{u}^2\big)-7 (\hat{s}+\hat{u}) \big(\hat{s}^2+\hat{u}^2\big)\nonumber\\
&&{}-8 \hat{t}^2 (\hat{s}+\hat{u})\big\}\,,
\ee
\begin{eqnarray}\label{gq3pd}
|\mathcal{M}^{\mathrm{Inc}(d)}[\leftidx{^{3}}{P}_{[J]}^{(8)}]|^2=-\frac{3}{5} \sum_J
|\mathcal{M}^{\mathrm{Inc}(f)}[\leftidx{^{3}}{P}_{[J]}^{(8)}]|^2\,.
\end{eqnarray}


\subsection{$g+\bar{q}\rightarrow J/\psi +\bar{q}$ channel}

The Feynman diagrams contributing to the $g+ \bar q \to J/\psi + \bar q$ channel can be obtained from those shown in Fig.~\ref{fig_gq} just reversing the quark line, while the corresponding color factors are collected in Table~\ref{table_gq} (lower signs).

The resulting amplitudes squared are the following:
\begin{eqnarray}\label{gqbar1s0f}
|\mathcal{M}^{\mathrm{Inc}(f)}[\leftidx{^{1}}{S}{_0}^{(8)}, \leftidx{^{3}}{S}{_1}^{(8)}, \leftidx{^{3}}{P}_J^{(8)}]|^2_{g\bar{q}} & = &
\mathcal{M}^{\mathrm{Inc}(f)}[\leftidx{^{1}}{S}{_0}^{(8)}, \leftidx{^{3}}{S}{_1}^{(8)}, \leftidx{^{3}}{P}_J^{(8)}]|^2_{gq}\,,\nonumber\\
|\mathcal{M}^{\mathrm{Inc}(d)}[\leftidx{^{1}}{S}{_0}^{(8)}, \leftidx{^{3}}{S}{_1}^{(8)}, \leftidx{^{3}}{P}_J^{(8)}]|^2_{g\bar{q}} & = & -
\mathcal{M}^{\mathrm{Inc}(d)}[\leftidx{^{1}}{S}{_0}^{(8)}, \leftidx{^{3}}{S}{_1}^{(8)}, \leftidx{^{3}}{P}_J^{(8)}]|^2_{gq}\,.
\end{eqnarray}

\subsection{$q+g\rightarrow J/\psi +q$ channel}
The color factors for the $q+g\rightarrow J/\psi +q$ channel are listed in Table~\ref{table_qg} (upper signs), and the corresponding amplitudes squared are:
\begin{table}[b]
  \centering
 \caption{Color factors of ${^3}{S}{_1^{(8)}}$, ${^1}{S}{_0^{(8)}}$ and  ${^3}{P}{_J^{(8)}}$
states for the $q+g\rightarrow J/\psi+q$ (upper signs) and $\bar q+g\rightarrow J/\psi+\bar q$ (lower signs) channels.}
  \label{table_qg}
   \begin{tabular}{cccccc}
   \toprule
  \hline
  \hline
  State  & Diagram &$C_U$ &$C_I$  &$C_F$  &$C^{\mathrm{ Inc}}$\\
 \hline
\midrule
$\leftidx{^3}{S}{_1^{(8)}}$
&\circled{A}$\times$\circled{A}&$\frac{N^2-1}{8 N^2}$ & $\mp\frac{1}{8}$ & $\mp\frac{1}{8\left(N^2-1\right)}$& $\mp\frac{N^2}{8\left(N^2-1\right)}$  \\
 $~~$ & \circled{A}$\times$\circled{B}&$-\frac{1}{8 N^2}$ & $\pm\frac{1}{8 \left(N^2-1\right)}$ &  $\mp\frac{1}{8 \left(N^2-1\right)}$&0 \\
 $~~$ &\circled{A}$\times$\circled{C}&$\pm\frac{1}{8}$ & $-\frac{N^2}{8\left(N^2-1\right)}$ & 0& $-\frac{N^2}{8\left(N^2-1\right)}$  \\

  \vspace{0.01cm}\\
   $~~$ & \circled{B}$\times$\circled{B}&$\frac{N^2-1}{8 N^2}$ & $\pm\frac{1}{8\left(N^2-1\right)}$ &$\pm\frac{1}{8}$&$\pm\frac{N^2}{8\left(N^2-1\right)}$ \\
    $~~$ & \circled{B}$\times$\circled{C}&$\mp\frac{1}{8}$ & 0 &$-\frac{N^2}{8\left(N^2-1\right)}$ &$-\frac{N^2}{8\left(N^2-1\right)}$\\

 \vspace{0.01cm}\\
  $~~$ &  \circled{C}$\times$\circled{C}&$\frac{1}{4}$ & $\mp\frac{N^2}{8\left(N^2-1\right)}$& $\pm\frac{N^2}{8\left(N^2-1\right)}$&0\\

 \vspace{0.01cm}\\
    $\leftidx{^1}{S}{_0^{(8)}}$\&  $\leftidx{^3}{P}{_J^{(8)}}$& \circled{C}$\times$\circled{C}&$\frac{N^2-4}{4N^2}$ & $\mp\frac{N^2-4}{8\left(N^2-1\right)}$ & $\pm\frac{N^2-4}{8\left(N^2-1\right)}$&0\\
  \hline
  \hline
\bottomrule
  \end{tabular}
\end{table}

\begin{eqnarray}\label{qg1s0}
|\mathcal{M}^{\mathrm{Inc}}[\leftidx{^{1}}{S}{_0}^{(8)}]|^2 = |\mathcal{M}^{\mathrm{Inc}}[\leftidx{^{3}}{P}_J^{(8)}]|^2 = 0\,,
\end{eqnarray}
\begin{equation}\label{gq3s1}
\begin{split}
|\mathcal{M}^{\mathrm{Inc}}[\leftidx{^{3}}{S}{_1}^{(8)}]|^2=&-\frac{3(4\pi\alpha_s)^3 }
{64M^3}\langle 0\mid \mathcal{O}_8^{J/\psi}(\leftidx{^3}{S}{_1})\mid 0\rangle\,
\frac{ (\hat{s}-\hat{t}) \left(\hat{s}^2+2 \hat{s} \hat{u}+2 \hat{u}^2+2 \hat{u} \hat{t}+\hat{t}^2\right)}{ \hat{s} \hat{t} (\hat{s}+\hat{t}) } \,.
\end{split}
\end{equation}
\subsection{$\bar{q}+g\rightarrow J/\psi +\bar{q}$ channel}
The color factors for the $\bar{q}+g\rightarrow J/\psi +\bar{q}$ channel are listed in Table~\ref{table_qg} (lower signs) and the corresponding amplitudes squared are:
\begin{eqnarray}\label{qbarg1s0}
|\mathcal{M}^{\mathrm{Inc}}[\leftidx{^{1}}{S}{_0}^{(8)}]|^2 & = & |\mathcal{M}^{\mathrm{Inc}}[\leftidx{^{3}}{P}_J^{(8)}]|^2=0\,,\\
\label{qbarg3s1}
|\mathcal{M}^{\mathrm{Inc}}[\leftidx{^{3}}{S}{_1}^{(8)}]|^2_{\bar{q}g} &= & -|\mathcal{M}^{\mathrm{Inc}}[\leftidx{^{3}}{S}{_1}^{(8)}]|^2_{qg}\,.
\end{eqnarray}

\subsection{$q+\bar{q}\rightarrow J/\psi +g$ channel}
\begin{figure}[h]
\begin{center}
\includegraphics[height=5cm,width=15cm]{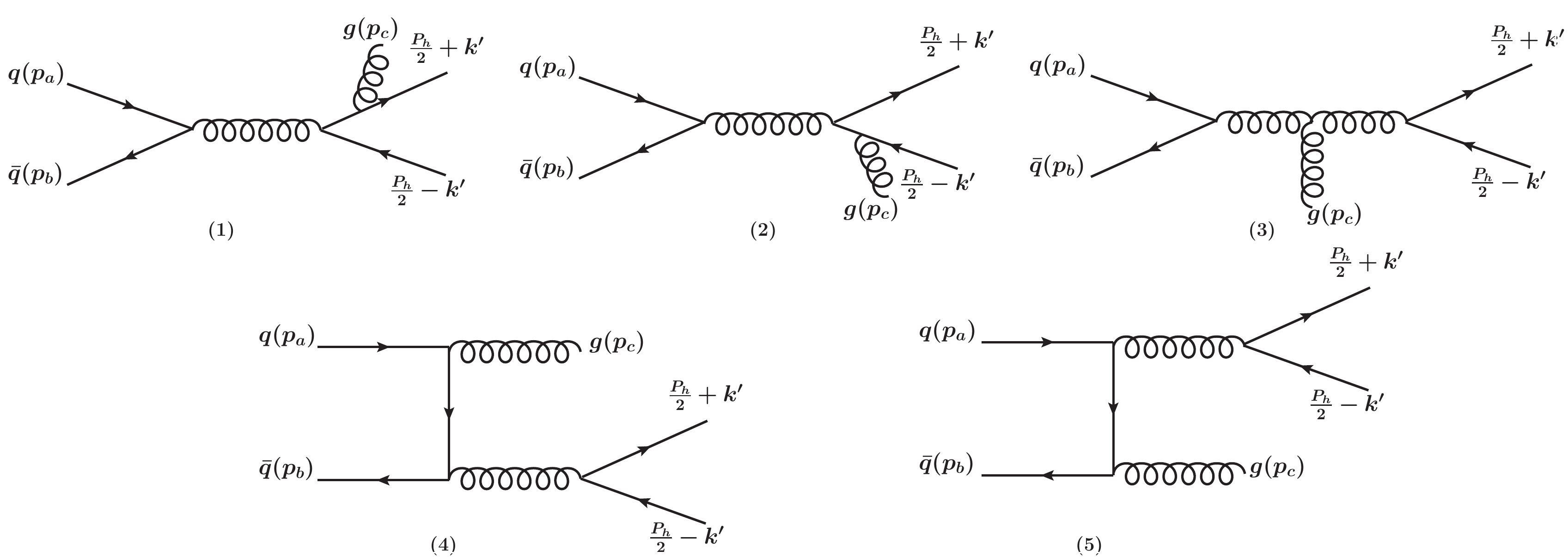}
\end{center}
\caption{\label{fig_qqbar} Feynman diagrams for the $q+\bar{q}\rightarrow J/\psi+g$ process.}
\end{figure}

The Feynman diagrams contributing to the $q + \bar q \to J/\psi + g$ channel are shown in Fig.~\ref{fig_qqbar}, while the corresponding color factors are collected in Table~\ref{table_qqbar} (column $q\bar q$ and upper signs).
Here the label \circled{A} represents the grouping of the $1^{\rm st}$, $2^{\rm nd}$ and $3^{\rm rd}$ Feynman diagrams, \circled{B} refers to the $4^{\rm th}$ diagram, while \circled{C} refers to the $5^{\rm th}$ diagram in Fig.~{\ref{fig_qqbar}}.
For $\leftidx{^1}{S}{_0^{(8)}}$ and $\leftidx{^3}{P}{_J^{(8)}}$ states, the $3^{\rm rd}$, $4^{\rm th}$ and $5^{\rm th}$ diagrams in Fig.~{\ref{fig_qqbar}} do not contribute.
\begin{table}[h!]
  \centering
 \caption{Color factors corresponding to the ${^3}{S}{_1^{(8)}}$, ${^1}{S}{_0^{(8)}}$ and ${^3}{P}{_J^{(8)}}$
states for the $q+\bar{q} \rightarrow J/\psi+g$ (column $q\bar q$, upper signs) and $\bar q+q \rightarrow J/\psi+g$ (column $\bar q q$, lower signs) channels.}
  \label{table_qqbar}
   \begin{tabular}{ccccccc}
   \toprule
  \hline
  \hline
  State &  $q\bar{q}$ Diagram & $\bar{q}q$ Diagram &$C_U$ &$C_I$ &$C_F$ &$C^{\mathrm{ Inc}}$\\
 \hline
\midrule
$\leftidx{^3}{S}{_1^{(8)}}$
&\circled{A}$\times$\circled{A}
&\circled{A}$\times$\circled{A}
&$\frac{N^2-1}{4N}$ & $\pm\frac{1}{4N}$ & $\pm\frac{N}{8}$ & $\pm\frac{N^2+2}{8 N}$   \vspace*{0.04cm}\\
  $~~$ &\circled{A}$\times$\circled{B}
  &\circled{A}$\times$\circled{C}
  &$\mp\frac{N^2-1}{8N}$ & $-\frac{1}{8N}$ &  0& $-\frac{1}{8N}$
  \vspace*{0.04cm}\\
   $~~$ &\circled{A}$\times$\circled{C}
 &\circled{A}$\times$\circled{B}
 &$\pm\frac{N^2-1}{8N}$ & $\frac{1}{8N}$ & $\frac{N}{8}$& $\frac{N^2+1}{8N}$   \\

  \vspace{0.01cm}\\
   $~~$ & \circled{B}$\times$\circled{B}
   & \circled{C}$\times$\circled{C}
   &$\frac{\left(N^2-1\right)^2}{8N^3}$ & $\mp\frac{1}{8N^3}$ &$\mp\frac{1}{8N}$ &$\mp\frac{N^2+1}{8N^3}$\vspace*{0.04cm}\\
    $~~$ & \circled{B}$\times$\circled{C}
    & \circled{B}$\times$\circled{C}
    &$-\frac{N^2-1}{8N^3}$ & $\mp\frac{N^2+1}{8N^3}$ &$\mp\frac{1}{8N}$ &$\mp\frac{2N^2+1}{8N^3}$\\

 \vspace{0.01cm}\\
 $~~$ &\circled{C}$\times$\circled{C}
 &\circled{B}$\times$\circled{B}
 &$\frac{\left(N^2-1\right)^2}{8N^3}$ & $\mp\frac{1}{8N^3}$& $\pm\frac{N^2-1}{8 N}$&$\pm\frac{N^2(N^2-1)-1}{8N^3}$\\

 \vspace{0.01cm}\\
    $\leftidx{^1}{S}{_0^{(8)}}$\&  $\leftidx{^3}{P}{_J^{(8)}}$& \circled{A}$\times$\circled{A}
    & \circled{A}$\times$\circled{A}
    &$\frac{(N^2-4)(N^2-1)}{4N^3}$ &  $\pm\frac{N^2-4}{4N^3}$ & $\pm\frac{N^2-4}{8N}$&$\pm\frac{(N^2-4)(N^2+2)}{8N^3}$\vspace*{0.04cm}\\
  \hline
  \hline
\bottomrule
  \end{tabular}
\end{table}


The resulting amplitudes squared are:
\begin{eqnarray}\label{qqbar1s0}
|\mathcal{M}^{\mathrm{Inc}}[\leftidx{^{1}}{S}{_0}^{(8)}]|^2_{q\bar{q}}=\frac{55(4\pi\alpha_s)^3}{432M}
\langle 0\mid \mathcal{O}_8^{J/\psi}(\leftidx{^1}{S}{_0})\mid 0\rangle\,
\frac{\hat{u}^2+\hat{t}^2}{\hat{s}(\hat{s}-M^2)^2}\,,
\end{eqnarray}

\begin{equation}\label{qqbar3s1}
\begin{split}
|\mathcal{M}^{\mathrm{Inc}}[\leftidx{^{3}}{S}{_1}^{(8)}]|^2_{q\bar{q}}=&\frac{(4\pi\alpha_s)^3}{648M^3}
\langle 0\mid \mathcal{O}_8^{J/\psi}(\leftidx{^3}{S}{_1})\mid 0\rangle
\frac{\left(71 \hat{u}^2-38 \hat{u} \hat{t}-10 \hat{t}^2\right) \left(2 \hat{s}^2+2 \hat{s} (\hat{u}+\hat{t})+\hat{u}^2+
\hat{t}^2\right)}{\hat{u} \hat{t} (\hat{u}+\hat{t})^2}\,,
\end{split}
\end{equation}

\begin{eqnarray}\label{qqbar3p0}
|\mathcal{M}^{\mathrm{Inc}}[\leftidx{^{3}}{P}{_0}^{(8)}]|^2_{q\bar{q}}=\frac{55(4\pi\alpha_s)^3}{324M^3}
\langle 0\mid \mathcal{O}_8^{J/\psi}(\leftidx{^3}{P}{_0})\mid 0\rangle\,
\frac{(\hat{s}-3M^2)^2(\hat{u}^2+\hat{t}^2)}{\hat{s}(\hat{s}-M^2)^4}\,,
\end{eqnarray}

\begin{eqnarray}\label{qqbar3p1}
|\mathcal{M}^{\mathrm{Inc}}[\leftidx{^{3}}{P}{_1}^{(8)}]|^2_{q\bar{q}}=\frac{55(4\pi\alpha_s)^3}{162M^3}
\langle 0\mid \mathcal{O}_8^{J/\psi}(\leftidx{^3}{P}{_1})\mid 0\rangle\,
\frac{4\hat{u}\hat{t}M^2+\hat{s}(\hat{u}^2+\hat{t}^2)}{(\hat{s}-M^2)^4}\,,
\end{eqnarray}

\begin{eqnarray}\label{qqbar3p2}
\begin{split}
|\mathcal{M}^{\mathrm{Inc}}[\leftidx{^{3}}{P}{_2}^{(8)}]|^2_{q\bar{q}} = {} & \frac{11(4\pi\alpha_s)^3}{162
   M^3 }
\langle 0\mid \mathcal{O}_8^{J/\psi}(\leftidx{^3}{P}{_2})\mid 0\rangle\,
\frac{(6 M^4+\hat {s}^2) (\hat{u}^2+\hat{t}^2)+12 M^2 \hat{s} \hat{u} \hat{t}}{\hat{s} (M^2-\hat{s})^4}\,.
\end{split}
\end{eqnarray}

By using Eq.~(\ref{eq:hqss}) we can sum the last three terms as
\begin{eqnarray}\label{qqbar3p}
|\mathcal{M}^{\mathrm{Inc}}[\leftidx{^{3}}{P}_{[J]}^{(8)}]|^2_{q\bar{q}}=\frac{55(4\pi\alpha_s)^3}{108M^3}
\langle 0\mid \mathcal{O}_8^{J/\psi}(\leftidx{^3}{P}{_0})\mid 0\rangle
\frac{(7 M^4 + 3\hat{s}^2) \big(\hat{u}^2+\hat{t}^2\big)-2 M^2 \hat{s} \big(\hat{u}^2-8 \hat{u} \hat{t}+\hat{t}^2\big)}{ \hat{s} \big(M^2-\hat{s}\big)^4}\,.
\end{eqnarray}

\subsection{$\bar{q}+q\rightarrow J/\psi +g$ channel}
The color factors for the $\bar{q}+q\rightarrow J/\psi +g$ channel are listed in Table~\ref{table_qqbar} (column $\bar q q$ and lower signs) and the corresponding amplitudes squared are:
\begin{eqnarray}\label{qbarq1s0}
|\mathcal{M}^{\mathrm{Inc}}[\leftidx{^{1}}{S}{_0}^{(8)},\leftidx{^{3}}{S}{_1}^{(8)}, \leftidx{^{3}}{P}_{J}^{(8)} ]|^2_{\bar{q}q}=-|\mathcal{M}^{\mathrm{Inc}}[\leftidx{^{1}}{S}{_0}^{(8)}, \leftidx{^{3}}{S}{_1}^{(8)}, \leftidx{^{3}}{P}_{J}^{(8)}]|^2_{q\bar{q}}\,.
\end{eqnarray}

\subsection{$g+g\rightarrow J/\psi$ channel}
\begin{figure}[!h]
\begin{center}
\includegraphics[height=3cm,width=13cm]{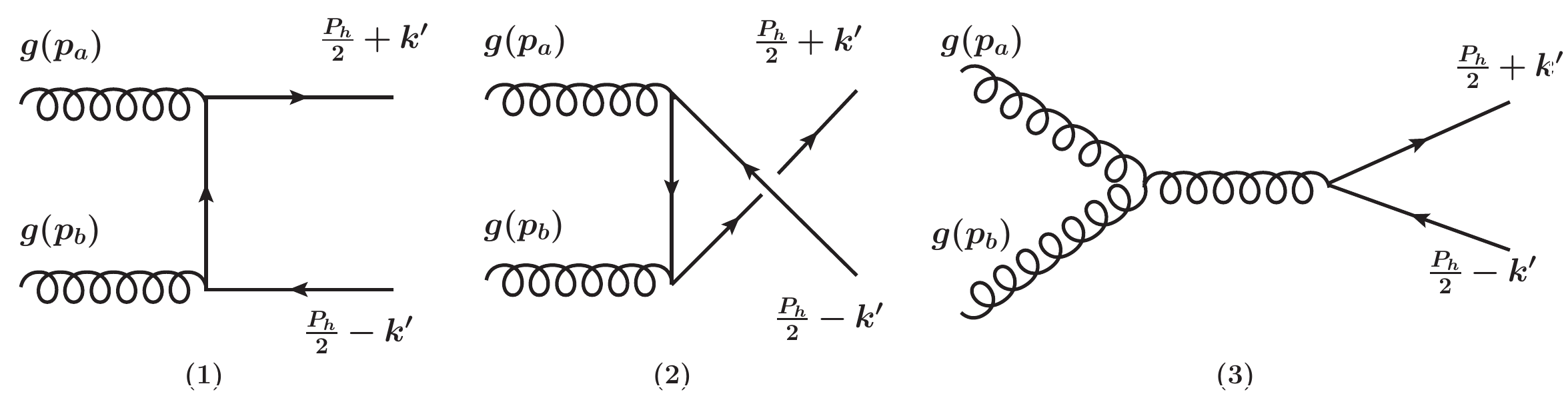}
\end{center}
\caption{\label{fig_gg2to1} Feynman diagrams for the $g+g\rightarrow J/\psi$ process.}
\end{figure}

The Feynman diagrams for the $g+g\rightarrow J/\psi$ process are shown in Fig.~\ref{fig_gg2to1}, while in Table~\ref{table_gg2to1} we give the color factors for the sum of them. The amplitude for the $\leftidx{^3}{S}{_1^{(8)}}$ state is identically zero.

 \begin{table}[!h]
  \centering
  \caption{Color factors of ${^1}{S}{_0^{(8)}}$  and ${^3}{P}{_J^{(8)}}$ states for the $g+g\rightarrow J/\psi $ channel.}
  \label{table_gg2to1}
   \begin{tabular}{cccc}
   \toprule
  \hline
  \hline
$C_U$ &$C_I^{(f)}$ &$C_F^{(f)}$ &$C^{\mathrm{ Inc}(f)}$ \\
 \hline
\midrule
   $\frac{N^2-4}{2N(N^2-1)}$ &
  $-\frac{N^2-4}{4N(N^2-1)}$ & $\frac{N^2-4}{4N(N^2-1)}$ &0\\
     \hline
  \hline
\bottomrule
  \end{tabular}
\end{table}

The other amplitudes squared, adopting the corresponding color factors, are:
\be
|\mathcal{M}^{\mathrm{Inc}(f,d)}[\leftidx{^{1}}{S}{_0}^{(8)}]|^2 = |\mathcal{M}^{\mathrm{Inc}(f,d)}[\leftidx{^{3}}{S}{_1}^{(1,8)}]|^2=
|\mathcal{M}^{\mathrm{Inc}(f,d)}[\leftidx{^{3}}{P}_J^{(8)}]|^2=0\,.
\ee

\subsection{$q+\bar{q}\rightarrow J/\psi$ and $\bar q + q \to J/\psi$ channels}
\begin{figure}[h]
\begin{center}
\includegraphics[height=3cm,width=6cm]{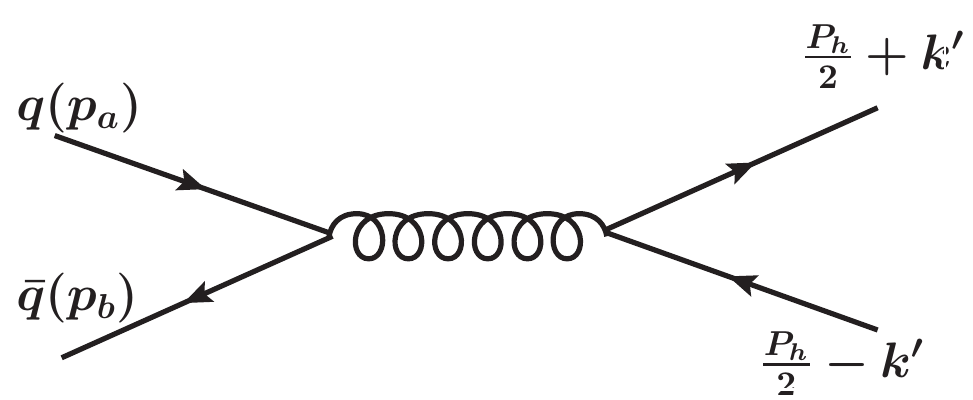}
\end{center}
\caption{\label{fig_qqbar2to1}Feynman diagram for the $q+\bar{q}\rightarrow J/\psi$ process.}
\end{figure}

The Feynman diagram for the $q (\bar q) +\bar{q} (q)\rightarrow J/\psi$ channel is shown in Fig.~\ref{fig_qqbar2to1}, while in Table~\ref{table_qqbar2to1} we give the corresponding color factors: $q\bar q$ (upper signs), $\bar q q$ (lower signs). Here only the $\leftidx{^3}S_1^{(8)}$ state contributes.

\begin{table}[h]
  \centering
  \caption{Color factors corresponding to the ${^3}{S}{_1^{(8)}}$  state for the $q+\bar{q}\rightarrow J/\psi $ (upper signs) and the $\bar q+q\rightarrow J/\psi$ (lower signs) channel.}
  \label{table_qqbar2to1}
   \begin{tabular}{cccc}
   \toprule
  \hline
  \hline
 $C_U$ &$C_I$ &$C_F$ &$C^{\mathrm{ Inc}}$\\
 \hline
\midrule\vspace*{0.04cm}
  $\frac{N^2-1}{4N^2}$ &
  $\pm\frac{1}{4N^2}$ & $\pm\frac{1}{4}$ &$\pm\frac{N^2+1}{4N^2}$\vspace*{0.04cm}\\
     \hline
  \hline
\bottomrule
  \end{tabular}
\end{table}
The amplitudes squared are:
\begin{eqnarray}\label{21qq3s1}
|\mathcal{M}^{\mathrm{Inc}}[\leftidx{^{3}}{S}{_1}^{(8)}]|^2_{q\bar{q}} = \frac{5(4\pi\alpha_s)^2}{108M}
\langle 0\mid \mathcal{O}_8^{J/\psi}(\leftidx{^3}{S}{_1})\mid 0\rangle\,,
\end{eqnarray}
\begin{eqnarray}\label{21qq3s11}
|\mathcal{M}^{\mathrm{Inc}}[\leftidx{^{3}}{S}{_1}^{(8)}]|^2_{\bar{q}q} = -|\mathcal{M}^{\mathrm{Inc}}[\leftidx{^{3}}{S}{_1}^{(8)}]|^2_{q\bar{q}}\,.
\end{eqnarray}

\subsection{Long Distance Matrix Elements}
%
In Table~\ref{table8} we collect the values of the LDMEs adopted in the present study:
\begin{table}[h]
  \centering
  \caption{Numerical values of LDMEs.}
  \label{table8}
   \begin{tabular}{ccccc}
    \toprule
  \hline
  \hline
  $~~~~~~~~~~~$
&$\langle  \mathcal{O}_1^{J/\psi}(\leftidx{^3}{S}{_1})\rangle$
&$~~\langle \mathcal{O}_8^{J/\psi}(\leftidx{^3}{S}{_1})\rangle$
&$~~\langle \mathcal{O}_8^{J/\psi}(\leftidx{^1}{S}{_0})\rangle$
 &$~~\langle \mathcal{O}_8^{J/\psi}(\leftidx{^3}{P}{_0})\rangle$  \\
 $~~~~~~~~~~~$
 &$\mathrm{GeV}^3$
&$~\times 10^{-2}\mathrm{GeV}^3$
&$~\times 10^{-2}\mathrm{GeV}^3$
&$~\times 10^{-2}\mathrm{GeV}^5$\\
 \hline
\midrule
BK11~\cite{Butenschoen:2011yh} &1.32 & $+0.224$ & $+4.97$   &$ -1.61$\\
SYY13~\cite{Sun:2012vc}        &     & $-0.93$  & $+14.23$ &$ -1.75\,m_c^2$\\
  \hline
  \hline
\bottomrule
  \end{tabular}
\end{table}

where we have used $m_c=M/2$.


\end{document}